\documentclass[10.5pt]{article}
\usepackage{amsfonts}
\usepackage{fullpage,graphicx,subfigure,mathdots,mathpazo,color}
\usepackage{amsmath,amscd,tikz,mathrsfs,cite}
\usepackage[normalem]{ulem}
\usepackage{amsmath}
\usepackage{setspace}
\usepackage{epsfig,amsmath,graphicx,amssymb,overpic}
\usepackage{bm}
\usepackage{graphicx}
\usepackage{subfigure, xcolor}
\usepackage{ntheorem}
\usepackage{listings,diagbox}
\usepackage{color}
\usepackage{epsfig,amsmath,graphicx,amssymb,overpic}
\usepackage{booktabs}
\usepackage{diagbox}
\usepackage{graphicx}
\usepackage{dcolumn}
\usepackage{bm}
\usepackage{graphicx}
\usepackage{subfigure}
\usepackage{epsfig,amsmath,graphicx,amssymb,overpic,cite}
\usepackage{makecell}
\usepackage{epsfig,amsmath,graphicx,amssymb,overpic}
\usepackage{bm}
\usepackage{graphicx}
\usepackage{subfigure, xcolor}
\usepackage{float}
\usepackage{listings}
\usepackage{tabularx}
\usepackage{multirow}
\usepackage{algorithm}
\usepackage{algorithmic}
\usepackage[section]{placeins}

\def\be{\begin{equation}}
\def\ee{\end{equation}}
\def\bee{\begin{eqnarray}}
\def\ene{\end{eqnarray}}
\def\bes{\begin{subequations}}
\def\ees{\end{subequations}}

\def\v{\vspace{0.1in}}

\def\be{\begin{equation}}
\def\ee{\end{equation}}
\def\bee{\begin{eqnarray}}
\def\ene{\end{eqnarray}}
\def\bes{\begin{subequations}}
\def\ees{\end{subequations}}

\def\d{\displaystyle}

\def\v{\vspace{0.1in}}

\newtheorem{remark}{Remark}

\begin{document}

\baselineskip=14pt \renewcommand {\thefootnote}{\dag}
\renewcommand
{\thefootnote}{\ddag} \renewcommand {\thefootnote}{ }

\pagestyle{plain}

\begin{center}
\baselineskip=16pt \leftline{} \vspace{-.3in} {\Large \textbf{%
Two-dimensional fractional discrete NLS equations: dispersion relations,
rogue waves, fundamental and vortex solitons}} \\[0.2in]

Ming Zhong$^{1,2}$,\,\, Boris A. Malomed$^{3,4}$,\,\, Jin Song$^{1,2}$,\,\,
Zhenya Yan$^{1,2,*}$\footnote{$^*$ Corresponding author at: KLMM, Academy of
Mathematics and Systems Science, Chinese Academy of Sciences, Beijing
100190, China. \textit{Email address}: zyyan@mmrc.iss.ac.cn (Corresponding
author)} \\[0.15in]
\textit{{\small $^1$KLMM, Academy of Mathematics and Systems Science,
Chinese Academy of Sciences, Beijing 100190, China\\[0pt]
$^2$School of Mathematical Sciences, University of Chinese Academy of
Sciences, Beijing 100049, China\\[0pt]
$^3$Department of Physical Electronics, School of Electrical Engineering,
Faculty of Engineering, Tel Aviv University, \\[0pt]
Tel Aviv 69978, Israel \vspace{-0.08in}\\[0pt]
$^4$Instituto de Alta Investigaci\'{o}n, Universidad de Tarapac\'{a},
Casilla 7D, Arica, Chile}} % (Date:\,\, \today)
\end{center}

%\noindent \rule[0.25\baselineskip]{\textwidth}{0.6pt}

%\begin{center}
%\vspace{0.1in} \noindent \textbf{Abstract}\thinspace\
%\end{center}

\vspace{0.1in} \noindent \textbf{Abstract}\,\, We introduce physically
relevant new models of two-dimensional (2D) fractional lattice media accounting
for the interplay of fractional intersite coupling and onsite self-focusing.
Our approach features novel discrete fractional operators based on an
appropriately modified definition of the continuous Riesz fractional derivative. The
model of the 2D isotropic lattice employs the discrete fractional Laplacian,
whereas the 2D anisotropic system incorporates discrete fractional
derivatives acting independently along orthogonal directions with different L%
\'{e}vy indices (LIs). We derive exact linear dispersion relations (DRs),
and identify spectral bands that permit linear modes to exist, finding them
to be similar to their continuous counterparts, apart from differences in
the wavenumber range. Additionally, the modulational instability in the
discrete models is studied in detail, and, akin to the linear DRs, it is
found to align with the situation in continuous models. This consistency
highlights the nature of our newly defined discrete fractional derivatives.
Furthermore, using Gaussian inputs, we produce a variety of rogue-wave
structures. By means of numerical methods, we systematically construct
families of 2D fundamental and vortex solitons, and examine their stability.
Fundamental solitons maintain the stability due to the discrete nature of
the interactions, preventing the onset of the critical and supercritical
collapse. On the other hand, vortex solitons are unstable in the isotropic
lattice model. However, the anisotropic one -- in particular, its symmetric
version with equal LIs acting in both directions -- maintains stable vortex
solitons with winding numbers $S=1$ and $S=3$. The detailed results stress
the robustness of the newly defined discrete fractional Laplacian in
supporting well-defined soliton modes in the 2D lattice media. %\end{f_4}
%\end{abstract}

\vspace{0.1in} \noindent \textbf{KEYWORDS} \thinspace\ 2D fractional
discrete nonlinear equations, discrete fractional Laplacian, modulation
instability, rogue waves, fundamental and vortex solitons, stability

\baselineskip=14pt

\section{Introduction}

In the course of the last thirty years, the fractional calculus has found
diverse realizations in physics \cite{b3,b4,b1}, including non-Gaussian
stochasticity \cite{fd1,nG,PR1,PR2}, quantum mechanics \cite%
{Lask2,Lask3,GuoXu,fd2,St13,Lask4}, optics \cite%
{fd-na,fd-np,Longhi,liu23,Zhang15,ZZ16,Shilong}, control theory \cite{b2},
Bose-Einstein condensates \cite{bec-fd}, charge transfer in solids \cite{b5}%
, etc. Fractional derivatives were first introduced as an abstract
mathematical concept~\cite{book1,book2,book3,fc-rev}, including definitions
such as the Riemann-Liouville and Caputo fractional derivatives \cite%
{caputo,Uchaikin}. The latter one, with a non-integer order $\alpha $, is
defined as%
\begin{equation}
D_{x}^{\alpha }u(x)=\frac{1}{\Gamma \left( 1-\{\alpha \}\right) }\int_{0}^{x}%
\frac{u^{\left( n\right) }(s)}{\left( x-s \right) ^{\left\{ \alpha \right\} }%
}ds,  \label{cap}
\end{equation}%
where $n\equiv \lbrack \alpha ]+1$ with $[\alpha ]$ and $\{\alpha \}\equiv
\alpha -[\alpha ]$ being the integer and fractional parts of $\alpha $,
respectively, $\Gamma(\cdot) $ is the Euler's Gamma-function, and $%
u^{(n)}(x)=d^nu(x)/dx^n$ denotes the usual derivative of integer-order $n $.

In physical applications, the relevant definition is a simpler one,\textit{\
viz}., the Riesz fractional derivative (RFD)~\cite{Riesz}, which follows the
intuitive idea that the fractional-order differentiation of wave function $%
u(x)$ in the coordinate space is represented by the multiplication by a
fractional power, $|k|^{\alpha}$, of wavenumber $k$ in the Fourier space:%
\begin{equation}
\left( \!-\frac{\partial ^{2}}{\partial x^{2}}\!\right) ^{\alpha /2}\!u(x)=%
\frac{1}{2\pi }\!\int_{-\infty }^{+\infty }dk|k|^{\alpha }\!\int_{-\infty
}^{+\infty }u(s)e^{ik(x-s)}ds,\qquad  \label{Riesz derivative}
\end{equation}%
where real $\alpha $, which usually takes values $1<\alpha \leq 2$, is
called the L\'{e}vy index (LI) \cite{Benoit}. Thus, RFD is not a
differential operator, but an integral (alias \textit{pseudo-differential})
one,  represented by the continuous  Riesz fractional integrals \cite{Klein-Stoilov}.
A well-known example of the implementation of RFD in physics is
provided by the Laskin's fractional Schr\"{o}dinger equation (FSE) for the
wave function of a particle whose stochastic motion in the classical regime
is performed by random \textit{L\'{e}vy flights}, with the average distance\
from the initial position, $x=0$, growing with time $t$ as $|x|\sim
t^{1/\alpha }$, where $\alpha $ is the same LI (Sometime, the LI $\alpha$
takes values $0<\alpha\leq 2$~\cite{Benoit}). In the scaled form, the
respective one- and two-dimensional (1D and 2D) FSEs in the fractional
quantum mechanics are \cite{Lask2,Lask3,Lask4},
\begin{eqnarray}
i\frac{\partial u}{\partial t} &=&\frac{1}{2}\left( -\frac{\partial ^{2}}{%
\partial x^{2}}\right) ^{\alpha /2}u+V(x)u,  \label{FSE} \\
i\frac{\partial u}{\partial t} &=&\frac{1}{2}\big(-\nabla^2\big) ^{\alpha
/2}u+V(x,y)u,  \label{2D-FSE}
\end{eqnarray}%
where $\nabla^2=\partial_x^2+\partial_y^2$ is the Laplacian,\, and the
fractional Laplacian $\big(-\nabla^2\big) ^{\alpha /2}$ can also be defined
similar to the 1D case (\ref{Riesz derivative}), $V(x)$ and $V\left(
x,y\right) $ are the respective real trapping potentials. The limit case of $%
\alpha =2$ corresponds to the usual integer-order Schr\"{o}dinger equations
in canonical quantum mechanics.

While fractional quantum mechanics based on Eqs. (\ref{FSE}) and (\ref%
{2D-FSE}) has not yet been realized experimentally, it was proposed by
Longhi \cite{Longhi} to emulate it in terms of the classical paraxial light
propagation in optical cavities, using the commonly known similarity between
the quantum-mechanical Schr\"{o}dinger equation and parabolic evolution
equation for the envelope of optical waves. In that context, the action of
RFD \eqref{Riesz derivative} may be realized by passing the
Fourier-decomposed light beam through a properly designed phase plate, while
the continuous equation appears as a result of averaging over many cycles of
the light circulation in the cavity. In 2023, Liu \textit{et al}~\cite{liu23}
reported the first experimental implementation of the fractional
group-velocity dispersion (GVD) in fiber lasers modeled by the generalized
FSE in the temporal domain,
\begin{equation}
i\frac{\partial u}{\partial z}=\left[ D_{\alpha }\left( -\frac{\partial ^{2}%
}{\partial \tau^{2}}\right) ^{\!\!\alpha /2}-\sum_{k=2,3,...}\frac{\gamma _{k}}{%
k!}\left( i\frac{\partial }{\partial \tau}\right) ^{k}\right] u+V(\tau)u,
\label{FSE-esp}
\end{equation}%
where $z$ is the propagation distance, $\tau$ the time variable, $D_{\alpha }$ a real
fractional-dispersion parameter, $\gamma _{k}$ the real $k$-th regular GVD
parameter, and $V(\tau)$ an effective potential.

The emulation of FSEs in terms of the cavity optics makes it possible to
essentially extend physically relevant model equations. In particular, it is
possible to replace the real effective potential, which represents the
refractive-index pattern in the cavity, by complex $\mathcal{PT}$-symmetric
ones~\cite{ZZ16}. Furthermore, the optical implementation suggests one to
add the usual cubic terms which represent the Kerr effect in optical media
\cite{KA}. The result is the fractional nonlinear Schr\"{o}dinger (NLS)
equations, such as
\begin{equation}
i\frac{\partial u}{\partial t}=\frac{1}{2}\big(-\nabla^2\big) ^{\alpha
/2}u+V(x,y)u+g|u|^{2}u  \label{FGPE}
\end{equation}%
in 2D. In this equation, a real nonlinearity coefficient $g=+1$ and $-1$
represents, respectively, self-defocusing and focusing effects, and $V(x,y)$
denotes an external potential, which is a real-valued function or a complex $%
\mathcal{PT}$-symmetric one, subject to constraint $V(-x,-y)=V^{\ast }(x,y)$%
, where $\ast $ stands for the complex conjugate. Fractional NLS equations 
with some external potentials or without potential
have been the subject of many theoretical works, which have predicted a
variety of fractional solitons, vortices, domain walls, and other nonlinear
states -- see, in particular, original works \cite%
{GuoXu,Huang08,Huang16,Zhang16,Yao18,Guo18,DW,Xie19,Zeng20,Li20,Qiu20,Zhong23,Zhong23prsa,Zhong23cp,Zhong24,Zan24,Feng-Su}
and reviews \cite{review,review2,rew3}. Solitons in fractional media with
the quadratic (second-harmonic-generating) nonlinearity have been predicted
too \cite{chi2}.

%{\color{red}The nonlocal discrete Schr\"odinger equation with dispersion terms featuring power-law decay and exponential decay has been extensively studied~\cite{PRE95}, in which several  stationary states were found. }
Theoretical works addressing fractional nonlinear media were
extended towards the consideration of their discrete counterparts (alias
fractional lattices) \cite%
{mathematical,Molina,Molina2,Molina-electro,fd3,Ki13}. In these works,
discrete counterparts of the fractional derivatives of the Riemann-Liouville
and Caputo types (see Eq.~(\ref{cap})) were introduced, which amount to
nonlocal couplings in the underlying lattices. Previously, similar
nonlocally coupled lattice dynamical models were introduced in other
contexts \cite{PRE95,CMP17}. However, the above-mentioned physical
realizations of fractional media suggest to introduce their discrete
counterparts corresponding to the lattice versions of RFD. More recently, with the aid of the newly-defined
discrete fractional derivative, a novel 1D model of this type was elaborated for the 1D fractional lattice with the
onsite cubic self-focusing nonlinearity~\cite{we}. Moreover, families of discrete
solitons of the single- and two-site types, produced by the model, were
constructed, and their stability and mobility were also explored~\cite{we}.

Continuous waves (CWs), alias plane waves, which are the simplest relevant
solutions of nonlinear wave equations, are characterized by the spatial
wavenumber, related to the respective temporal frequency~\cite%
{MI0,MI1,MI2,MI3}. A key feature of CWs in dispersive wave media is their
modulation instability (MI, alias the Benjamin-Feir instability~\cite{MI0}),
which makes the CWs unstable under certain conditions~\cite{MI3}. In
particular, it is well known that both continuous and discrete NLS equations
give rise to MI in the case of the self-focusing nonlinearity. MI is a
significant topic across many areas, including fluid dynamics~\cite{MI0,Wh65}%
, nonlinear optics~\cite{Be66,Os67}, and plasmas~\cite{Wa68,Ha70}.
%Despite its extensive history spanning five decades, research on MI continues to be vibrant and ongoing. For example, recent studies have applied MI in the dynamics of Bose-Einstein condensates (BECs) to generate bright solitons in attractive condensates~\cite{St02} and to create Faraday waves in repulsive condensates~\cite{Ni07}.
In particular, rogue waves (RWs) are a well-known kind of wave phenomena
related to MI, which draws growing interest in various fields, such as
nonlinear optics~\cite{So07,Kwok}, deep ocean~\cite{orw,Ch11}, superfluids~%
\cite{superfluid}, plasma physics \cite{RW_plasma}, Bose-Einstein
condensates~\cite{bec-rw,bec-rw2}, atmosphere~\cite{Iafrati}, and even
financial markets~\cite{yanfrw,yanfrw2}. RWs are large-amplitude
spontaneously generated nonlinear waves that appear suddenly and disappear
just as quickly~\cite{Ak09}. In 1983, the exact first-order RW (alias the
Peregrine soliton/rogon) was found by Peregrine as an exact solution of the
integrable focusing NLS equation~\cite{Pe83}. Then, it was shown that the
Peregrine solitons explain diverse numerical and experimental results~\cite%
{Ki10}. In addition to the fact that the continuous nonlinear wave equations
produce RWs (see e.g., Refs.~\cite{RW-book,RW-rew1,RW-rew2,RW-rew3,RW-rew4}
and references therein), some discrete integrable nonlinear systems also
admit RW solutions~\cite{An10, An13, Yan12, Oh14, We18, Fe21, Ch14,
We16,Chen24}. Recently, RWs were found in the continuous two-L\'{e}vy-index
fractional Kerr media~\cite{Zhong24}. However, the study of RWs was not yet
developed in detail in terms of fractional discrete non-integrable systems.
With regard to the presence of MI in these systems, we here explore RWs as a
linear superposition of CWs and Gaussian perturbations with different
parameters in the fractional discrete systems.

The general objective of the present work is to extend the formulation and
analysis of 1D fractional discrete systems for 2D lattices with the onsite
self-focusing nonlinearity. A new straightforward possibility, offered by
the consideration of the 2D setting, is to construct 2D vortex solitons, in
addition to the fundamental lattice ones. Two different 2D fractional
discrete models are introduced in Section 2. One model is isotropic, with
the discrete version of the fractional Laplacian, see Eq.~(\ref{2D}) below.
The other model is, generally speaking, anisotropic, with two quasi-1D
discrete fractional derivatives acting separately along the two directions
of the underlying lattice, each derivative being defined by its own
coefficient and LI value, see Eq.~(\ref{FNLSE2}) below. A special role is
played by the symmetric version of the latter model, with equal coefficients
and LIs of both fractional derivatives.
%Additionally, we investigate the linear dispersion relations (DR) and MI phenomenon, which can provide the potential for the existence of high-amplitude nonlinear waves. The fractional discrete nonlinear Schr\"{o}dinger (DNLS) equations of both types are introduced in Section 2.
We derive exact linear dispersion relations (DRs) for these 2D fractional
lattice models, and compare them with the known result for the classical
discrete NLS (DNLS) equation in Section 3. The MI and RW generation are
investigated in Section 4, showing similarities to the results for
continuous models. Systematically collected results for the structure and
stability of 2D lattice solitons of the fundamental and vortex types,
produced by means of numerical methods, are reported, respectively, in
Sections 5. In particular, the discreteness provides stability of the 2D
solitons against the critical and supercritical collapse in the cases of LI $%
=2$ and LI $<2$, respectively. Vortex solitons are unstable in the\
framework of the isotropic model with the fractional Laplacian, while the
model with the independent quasi-1D fractional derivatives produce stable
vortex derivatives, with winding numbers (topological charges) $S=1$ and $3$%
. The paper is completed by Section 6.

\section{2D fractional DNLS equations with quasi-Riesz fractional derivatives
(RFDs)}

%\subsection {One-dimension Case}
%A complex function $u_{n}$ of the discrete coordinate $n=0,\pm 1,\pm 2,\pm
%3,...$, can be represented in\ terms of its Fourier transform, $U(k)$, which
%is defined in interval%
%\begin{equation}
%-\pi <k<+\pi ,  \label{pipi1}
%\end{equation}%
%as a periodic function of the real continuous coordinate $k$ in the Fourier
%space. The direct and inverse relations between $u_{n}$ and $U(k)$ are%
%\begin{equation}\label{inverse1}
%u_{n}=\frac{1}{2\pi }\int_{-\pi }^{+\pi }e^{-ikn}U(k)dk, \quad
%U(k)=\sum_{n=-\infty }^{+\infty }u_{n}e^{ikn}.
%\end{equation}
%
%The direct counterpart of the Riesz fractional derivative, with the Levy
%index $\alpha $, can be derived as follows:%
%\begin{gather}
%\left( -\frac{\partial ^{2}}{\partial n^{2}}\right) ^{\alpha /2}u_{n}=\frac{1%
%}{2\pi }\int_{-\pi }^{+\pi }e^{-ikn}|k|^{\alpha }dk\sum_{m=-\infty
%}^{+\infty }u_{m}e^{ikm}  \notag \\
%\equiv \frac{1}{\pi }\sum_{m=-\infty }^{+\infty }u_{m}\int_{0}^{\pi }\cos
%\left( (m-n)k\right) \cdot k^{\alpha }dk  \label{Riesz1} \\
%\equiv \sum_{l=-\infty }^{+\infty }D_{l}^{(\alpha )}u_{n+l},  \notag
%\end{gather}%
%where the coupling coefficients are%
%\begin{equation}
%D_{l}^{(\alpha )}\equiv \frac{1}{\pi }\int_{0}^{\pi }\cos \left( lk\right)
%\cdot k^{\alpha }dk.  \label{C1}
%\end{equation}%

%\subsection {Two-dimension Case}

\subsection{Definition of 2D discrete fractional derivatives}

A complex function $u_{n,m}=u(n,m,\cdot )$ of two discrete integer-value
coordinates $\left( n,m\right) $ and other variables can be represented by
its Fourier transform, $U_{k_{x},k_{y}}=U(k_{x},k_{y},\cdot )$, which is
defined in the interval of $k_{x},k_{y}\in \lbrack -\pi ,+\pi ]$, as a
periodic function of real continuously varying wavenumbers $\left(
k_{x},\,k_{y}\right) $ in the 2D Fourier space. The direct and inverse
Fourier relations between $u_{n,m}$ and $U_{k_{x},k_{y}}$ take the usual
form,%
\begin{equation}
\begin{array}{l}
u_{n,m}=\displaystyle\frac{1}{4\pi ^{2}}\int_{-\pi }^{+\pi }\int_{-\pi
}^{+\pi }e^{i\left( k_{x}n+k_{y}m\right) }U_{k_{x},k_{y}}dk,\vspace{0.1in}
\\
U_{k_{x},k_{y}}=\displaystyle\sum_{n,m=-\infty }^{+\infty
}u_{n,m}e^{-i\left( k_{x}n+k_{y}m\right) }.%
\end{array}%
\end{equation}
Thus, the new discrete counterpart of the 2D RFD (fractional Laplacian), $\left(
-\widehat{\partial }^{2}/\widehat{\partial }n^{2}-\widehat{\partial }^{2}/%
\widehat{\partial }m^{2}\right) ^{\alpha /2}$, with the LI $\alpha \in (1,2]$%
, is defined starting from its natural definition in the Fourier space:%
%\begin{widetext}
\begin{equation}  \label{2dd}
\begin{array}{rl}
\displaystyle\left\{ \!\!\left( \!\!-\frac{\widehat{\partial }^{2}}{\widehat{%
\partial }n^{2}}\!-\!\frac{\widehat{\partial }^{2}}{\widehat{\partial }m^{2}}%
\!\!\right) ^{\alpha /2}\!u\!\right\} _{n,m}= & \displaystyle\!\!\!\!\frac{1%
}{4\pi ^{2}}\int_{-\pi }^{+\pi }\!\!\int_{-\pi }^{+\pi }e^{i\left(
k_{x}n+k_{y}m\right) }\left( k_{x}^{2}+k_{y}^{2}\right) ^{\alpha
/2}dk_{x}dk_{y}\sum_{p,q=-\infty }^{+\infty }u_{p,q}e^{-i\left(
k_{x}p+k_{y}q\right) }\vspace{0.1in} \\
= & \displaystyle\!\!\!\!\frac{1}{4\pi ^{2}}\sum_{p,q=-\infty }^{+\infty
}u_{p,q}\int_{-\pi }^{+\pi }\!\!\int_{-\pi }^{+\pi }\cos \!\left(
\!k_{x}(p\!-\!n)\!+\!k_{y}(q\!-\!m)\!\right) \!\!\left(
k_{x}^{2}\!+\!k_{y}^{2}\right) ^{\alpha /2}dk_{x}dk_{y}\vspace{0.1in} \\
\equiv & \displaystyle\!\!\!\!\sum_{l_{x},l_{y}=-\infty }^{+\infty
}D_{l_{x},l_{y}}^{(\alpha )}u_{n+l_{x},m+l_{y}},%
\end{array}%
\end{equation}%
where the caret symbol ($\symbol{94}$) indicates the discrete character of
the operator, notation $\left\{ {}\right\} _{n,m}$ implies that it is the
value of the fractional Laplacian at the site with coordinates $\left(
n,m\right) $, and the coupling coefficients, which are even functions of $%
l_{x,y}$, are defined as%
\begin{equation}
D_{l_{x},l_{y}}^{(\alpha )}=\frac{1}{4\pi ^{2}}\int_{-\pi }^{+\pi
}\int_{-\pi }^{+\pi }\cos \left( k_{x}l_{x}\right) \cos \left(
k_{y}l_{y}\right) \left( k_{x}^{2}+k_{y}^{2}\right) ^{\alpha /2}dk_{x}dk_{y}.
\label{C1}
\end{equation}%
Note that, in the limit of $\alpha =2$ (the non-fractional case),
coefficients (\ref{C1}) are different from zero only in the vertical and
horizontal directions, i.e., for $l_{x}=0$ or $l_{y}=0$; for instance, $%
D_{l_{x}\neq 0,l_{y}=0}^{(\alpha )}=2(-1)^{l_{x}}l_{x}^{-2}$, which exactly
coincide with the coupling coefficients in the 1D fractional DNLS equation
\cite{we}. In 1D, the limit of RFD corresponding to $\alpha =2$ may
represent a physical system built as an array of quasi-1D Bose-Einstein
condensates (narrow tubes) of dipolar atoms \cite{we}. In 2D, it is possible
to consider a similar network of parallel quasi-1D condensates, with the
transverse cross section shaped as a square lattice, but this interpretation
requires special analysis which will be reported elsewhere.

Similarly, we can also define the RFD of function\ $u_{n,m}$ with respect to
the single direction in the 2D space, %\begin{widetext}
\begin{equation}
\begin{array}{rl}
\displaystyle\left\{ \left( -\frac{\widehat{\partial }^{2}}{\widehat{%
\partial }n^{2}}\right) ^{\alpha /2}u\right\} _{n,m}= & \displaystyle\frac{1%
}{4\pi ^{2}}\int_{-\pi }^{+\pi }\int_{-\pi }^{+\pi }e^{-i\left(
k_{x}n+k_{y}m\right) }|k_{x}|^{\alpha }dk_{x}dk_{y}\sum_{p,q=-\infty
}^{+\infty }u_{p,q}e^{i\left( k_{x}p+k_{y}q\right) }\vspace{0.1in} \\
= & \displaystyle\frac{1}{4\pi ^{2}}\sum_{p,q=-\infty }^{+\infty
}u_{p,q}\int_{-\pi }^{+\pi }\int_{-\pi }^{+\pi }\cos \left(
k_{x}(p-n)+k_{y}(q-m)\right) |k_{x}|^{\alpha }dk_{x}dk_{y}\vspace{0.1in} \\
\equiv & \displaystyle\sum_{l_{x},l_{y}=-\infty }^{+\infty
}E_{l_{x},l_{y}}^{(\alpha )}u_{n+l_{x},m+l_{y}},%
\end{array}
\label{Riesz3}
\end{equation}
where the real coupling coefficients are
\begin{equation}
E_{l_{x},l_{y}}^{(\alpha )}=\!\left\{ \!%
\begin{array}{cc}
0, & l_{y}\neq 0,\vspace{0.1in} \\
\dfrac{1}{\pi }\displaystyle\int_{0}^{\pi }\cos \left( k_{x}l_{x}\right)
k_{x}^{\alpha }dk_{x}\equiv E_{l_{x}}^{(\alpha )}, & l_{y}=0.%
\end{array}%
\right.  \label{C2}
\end{equation}%
In fact, the same operator as given by Eqs. (\ref{Riesz3}) and (\ref{C2})
appears in the respective 1D model:
\begin{equation}
\begin{array}{rl}
\left\{ \left( -\dfrac{\widehat{\partial }^{2}}{\widehat{\partial }n^{2}}%
\right) ^{\alpha /2}u\right\} _{n,m}= & \displaystyle\frac{1}{2\pi }%
\int_{-\pi }^{+\pi }e^{-ik_{x}n}|k_{x}|^{\alpha }dk_{x}\sum_{p=-\infty
}^{+\infty }u_{p}e^{ik_{x}p}\vspace{0.1in} \\
= & \displaystyle\frac{1}{2\pi }\sum_{p=-\infty }^{+\infty }u_{p}\int_{-\pi
}^{+\pi }\cos \left( k_{x}(p-n)\right) |k_{x}|^{\alpha }dk_{x}\vspace{0.1in}
\\
\equiv & \displaystyle\sum_{l_{x}=-\infty }^{+\infty }E_{l_{x}}^{(\alpha
)}u_{n+l_{x},m},%
\end{array}%
\end{equation}%
with the same coupling coefficients $E_{l_{x}}^{(\alpha )}$ as defined in
Eq. (\ref{C2}) \cite{we}{.}

\subsection{2D fractional DNLS equations}

With these definitions of discrete fractional derivatives given by Eqs.~%
\eqref{2dd} and \eqref{Riesz3}, the 2D isotropic fractional DNLS equation in
the dimensionless form is written as (in the optics notation, with the
evolution variable defined as the propagation distance $z$, cf. Eq. (\ref%
{FSE-esp}))
\begin{equation}
i\frac{du_{n,m}}{dz}=C\left\{ \left( -\frac{\widehat{\partial }^{2}}{%
\widehat{\partial }n^{2}}-\frac{\widehat{\partial }^{2}}{\widehat{\partial }%
m^{2}}\right) ^{\alpha /2}u\right\} _{n,m}-g|u_{n,m}|^{2}u_{n,m},  \label{2D}
\end{equation}%
Further, the 2D anisotropic fractional DNLS equation, with independent
fractional derivatives, characterized by the respective LIs $\alpha $ and $%
\beta $\, ($\alpha,\, \beta\in (1, 2]$), acting in two directions, is
introduced as
\begin{equation}
i\frac{du_{n,m}}{dz}=\left\{ \left[ C_{\alpha }\left( -\frac{\widehat{%
\partial }^{2}}{\widehat{\partial }n^{2}}\right) ^{\alpha /2}+C_{\beta
}\left( -\frac{\widehat{\partial }^{2}}{\widehat{\partial }m^{2}}\right)
^{\beta /2}\right] u\right\} _{n,m}-g|u_{n,m}|^{2}u_{n,m}\text{,}
\label{FNLSE2}
\end{equation}%
where the positive parameters $C,\,C_{\alpha },C_{\beta }$ are the
coefficients of the fractional discrete diffraction, alias the linear
coupling strength between adjacent sites of the lattice, and $g=+1$ and $-1$
represents, respectively, self-focusing and defocusing effects.

Equations~(\ref{2D}) and (\ref{FNLSE2}) conserve their Hamiltonians which
are, respectively,
\begin{equation}
H_{1}=C\sum_{m,n,p,q}D_{n-p,m-q}^{(\alpha )}u_{n,m}^{\ast }u_{p,q}-\frac{g}{2%
}\sum_{m,n}|u_{n,m}|^{4},  \label{Ha}
\end{equation}%
and
\begin{equation}
H_{2}=C_{\alpha }\sum_{n,p,m}E_{n-p}^{(\alpha )}u_{n,m}^{\ast
}u_{p,m}+C_{\beta }\sum_{n,m,q}E_{m-q}^{(\beta )}u_{n,m}^{\ast }u_{n,q}-%
\frac{g}{2}\sum_{m,n}|u_{m,n}|^{4}.  \label{H2}
\end{equation}%
We mainly consider the case of $C_{\alpha }=C_{\beta }\equiv C$ in Eq.~(\ref%
{FNLSE2}), with the anisotropy primarily represented by the different LIs.
Both Eqs.~(\ref{2D}) and (\ref{FNLSE2}) also conserve the total power
(norm), defined by the obvious expression:
\begin{equation}
P=\sum_{m.n}|u_{n,m}|^{2}.  \label{Po}
\end{equation}

\begin{remark}
Note that these fractional DNLS equations given by Eqs.~(\ref{2D}) and (\ref{FNLSE2}) can
also include a real or complex $\mathcal{PT}$-symmetric external potential $%
V(m,n,z)$, and the discrete cubic nonlinear term may be replaced by other
ones~\cite{we,Le08,Ke09}, such as the cubic-quintic competing nonlinearity, $%
g_{1}|u_{n,m}|^{2}u_{n,m}+g_{2}|u_{n,m}|^{4}u_{n,m}$, the power-law term, $%
|u_{n,m}|^{2p}u_{n,m}$, the saturable one, $%
|u_{n,m}|^{2}u_{n,m}/(1+S|u_{n,m}|^{2})$, and etc.. The 2D fractional DNLS equation can also be extended to the coupled cases, for example, the
2D isotropic coupled fractional DNLS equations
\begin{equation}
\begin{array}{l}
\d i\frac{du_{n,m}}{dz}=C\left\{ \left( -\frac{\widehat{\partial }^{2}}{%
\widehat{\partial }n^{2}}-\frac{\widehat{\partial }^{2}}{\widehat{\partial }%
m^{2}}\right) ^{\alpha /2}u\right\} _{n,m}+V_1(n,m,z)u_{n,m}-(g_{11}|u_{n,m}|^{2}+g_{12}|v_{n,m}|^{2})u_{n,m}, \\[2em]
\d i\frac{dv_{n,m}}{dz}=C\left\{ \left( -\frac{\widehat{\partial }^{2}}{%
\widehat{\partial }n^{2}}-\frac{\widehat{\partial }^{2}}{\widehat{\partial }%
m^{2}}\right) ^{\alpha /2}v\right\} _{n,m}+V_2(n,m,z)v_{n,m}-(g_{12}|u_{n,m}|^{2}+g_{22}|v_{n,m}|^{2})v_{n,m},
\end{array}
\end{equation}%
and 2D anisotropic coupled fractional DNLS equations
\begin{equation}
\begin{array}{l}
\d i\frac{du_{n,m}}{dz}=\left\{ \left[ C_{\alpha }\left( -\frac{\widehat{%
\partial }^{2}}{\widehat{\partial }n^{2}}\right) ^{\alpha /2}+C_{\beta
}\left( -\frac{\widehat{\partial }^{2}}{\widehat{\partial }m^{2}}\right)
^{\beta /2}\right] u\right\} _{n,m}+V_1(n,m,z)u_{n,m}-(g_{11}|u_{n,m}|^{2}+g_{12}|v_{n,m}|^{2})u_{n,m}, \\[2em]
\d i\frac{dv_{n,m}}{dz}=\left\{ \left[ C_{\alpha }\left( -\frac{\widehat{%
\partial }^{2}}{\widehat{\partial }n^{2}}\right) ^{\alpha /2}+C_{\beta
}\left( -\frac{\widehat{\partial }^{2}}{\widehat{\partial }m^{2}}\right)
^{\beta /2}\right] v\right\} _{n,m}+V_2(n,m,z)v_{n,m}-(g_{12}|u_{n,m}|^{2}+g_{22}|v_{n,m}|^{2})v_{n,m}.
\end{array}
\end{equation}%
Of course, these 2D fractional DNLS equations can also be extended to the 3D case.
\end{remark}

\begin{remark}
When $C_{\alpha }=C_{\beta }=C,\,\alpha =\beta $, Eq.~\eqref{FNLSE2} reduces
to
\begin{equation}
i\frac{du_{n,m}}{dz}=C\left\{ \left[ \left( -\frac{\widehat{\partial }^{2}}{%
\widehat{\partial }n^{2}}\right) ^{\alpha /2}+\left( -\frac{\widehat{%
\partial }^{2}}{\widehat{\partial }m^{2}}\right) ^{\alpha /2}\right]
u\right\} _{n,m}-g|u_{n,m}|^{2}u_{n,m}\text{,}  \label{FNLSE2-r}
\end{equation}%
It follows from the definition of the discrete fractional derivative that  quasi-1D model Eq.~\eqref{FNLSE2-r} with same LI is different
from Eq.~\eqref{2D} with the single LI.
\end{remark}

\begin{remark}
The following relations hold for the discrete fractional derivatives of the
CW:
\begin{equation}  \label{proof}
\begin{aligned} \left\{ \left( -\frac{\widehat{\partial
}^{2}}{\widehat{\partial }n^{2}} -\frac{\widehat{\partial
}^{2}}{\widehat{\partial}m^{2}}\right)^{\alpha/2}e^{i\textbf{k}\cdot
\textbf{r}}\right\} _{n,m}&=|\textbf{k}|^{\alpha}e^{i\textbf{k}\cdot
\textbf{r}},\\ \left\{ \left( -\frac{\widehat{\partial
}^{2}}{\widehat{\partial }n^{2}}\right)^{\alpha/2}e^{i\textbf{k}\cdot
\textbf{r}}\right\} _{n,m}&=|k_x|^{\alpha}e^{i\textbf{k}\cdot \textbf{r}},
\end{aligned}
\end{equation}%
with $\mathbf{k}=(k_{x},k_{y})$ and $\mathbf{r}=(n,m)$. They can be
substantiated through a straightforward Fourier expansion, and are utilized
below.  The proof of these relations is presented in Appendix A.
\end{remark}

\begin{remark}
In the limits of LIs $\alpha =\beta =1$, which is opposite to the one
corresponding to the non-fractional diffraction ($\alpha =\beta =2$), the
fractional Laplacians with exponents $\alpha /2$ or/ and $\beta /2$ in Eqs.~(%
\ref{2D}) and (\ref{FNLSE2}) reduce to the relativistic operators widely
used in mathematical physics~\cite{Longhi,RO,RO2}. In this case, Eqs.~(\ref%
{2D}) and (\ref{FNLSE2}) become
\begin{equation}
i\frac{du_{n,m}}{dz}=C\left\{ \left( -\frac{\widehat{\partial }^{2}}{%
\widehat{\partial }n^{2}}-\frac{\widehat{\partial }^{2}}{\widehat{\partial }%
m^{2}}\right) ^{1/2}u\right\} _{n,m}-g|u_{n,m}|^{2}u_{n,m},  \label{2Dr}
\end{equation}%
and
\begin{equation}
i\frac{du_{n,m}}{dz}=\left\{ \left[ C_{\alpha }\left( -\frac{\widehat{%
\partial }^{2}}{\widehat{\partial }n^{2}}\right) ^{1/2}+C_{\beta }\left( -%
\frac{\widehat{\partial }^{2}}{\widehat{\partial }m^{2}}\right) ^{1/2}\right]
u\right\} _{n,m}-g|u_{n,m}|^{2}u_{n,m}\text{.}  \label{FNLSE2r}
\end{equation}
\end{remark}

\begin{figure}[t]
\centering
\vspace{-0.15in} {\scalebox{0.67}[0.67]{\includegraphics{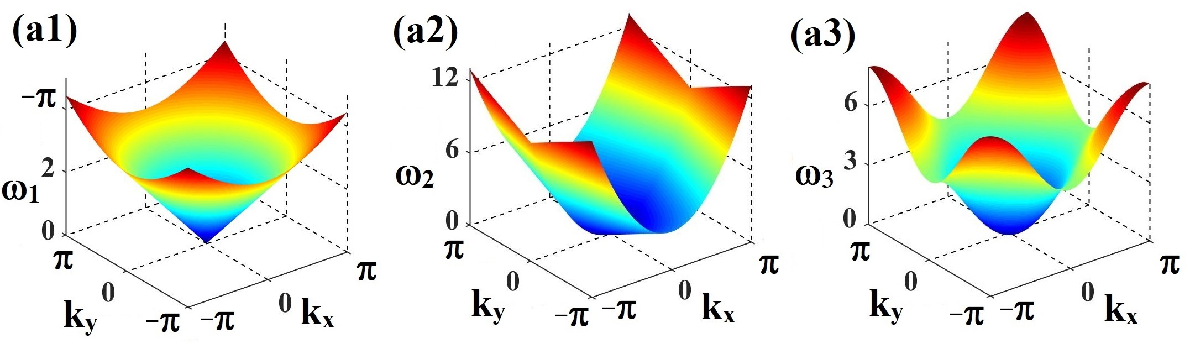}}}\hspace{-0.3in}
\vspace{-0.1in}
\caption{Different types of linear DRs: for Eq.~\eqref{2D} it is $\protect%
\omega _{1}(k)$ with LI $\protect\alpha =1$ (a1), for Eq.~\eqref{FNLSE2} it
is $\protect\omega _{2}(k)$ with LIs $\protect\alpha =2,\protect\beta =1$
(a2) , and for Eq.~(\protect\ref{DNLS}) it is $\protect\omega _{3}(k)$ (a3).
The coupling coefficients are $C=C_{\protect\alpha }=C_{\protect\beta }=1$.}
\label{LR}
\end{figure}

%  Predictions of the linear stability results are then verified by direct simulations of the perturbed evolution.

\section{Dispersion relations (DRs) and spectral band for linear modes}

An essential attribute of discrete systems is encapsulated in its DR,
delineating the connection between real frequency $\omega(\mathbf{k}) $ and
wavenumber $\mathbf{k}$ of small-amplitude lattice waves, commonly referred
to as \textquotedblleft phonons". Determined by the linearized system, the
DR also defines the spectral band, which is the range of frequencies that
permit the propagation of these linear waves \cite{Ke09}.

To establish the DRs, we consider a solution of the linearized version of
Eqs.~(\ref{2D}) and (\ref{FNLSE2}) in the CW form,
\begin{eqnarray}
u_{n,m}=\exp (i(\mathbf{k}\cdot \mathbf{r}-\omega_{j}(\mathbf{k})z)),\qquad
j=1,2,
\end{eqnarray}
respectively. This formulation leads to corresponding DRs in the form of
\begin{equation}
\omega _{1}(\mathbf{k})=C\sum_{l_{x}=-\infty }^{+\infty }\sum_{l_{y}=-\infty
}^{+\infty }D_{l_{x},l_{y}}^{(\alpha )}\cos \left(
l_{x}k_{x}+l_{y}k_{y}\right)  \label{om}
\end{equation}%
for Eq.~(\ref{2D}) as well as
\begin{equation}
\omega _{2}(\mathbf{k})=C\left[ \sum_{l_{x}=-\infty }^{+\infty
}E_{l_{x}}^{(\alpha )}\cos \left( l_{x}k_{x}\right) +\sum_{l_{y}=-\infty
}^{+\infty }E_{l_{y}}^{(\beta )}\cos \left( l_{y}k_{y}\right) \right]
\label{om2}
\end{equation}
for Eq.~(\ref{FNLSE2}). A straightforward application of Fourier integrals
reveals that the aforementioned expressions can be simplified to:
\begin{equation}
\begin{array}{l}
\omega _{1}(\mathbf{k})=C|\mathbf{k}|^{\alpha },\vspace{0.1in} \\
\omega _{2}(\mathbf{k})=C(|k_{x}|^{\alpha }+|k_{y}|^{\beta }),%
\end{array}
\label{oms}
\end{equation}%
see Appendix A for their detailed derivations.

\begin{figure}[!t]
\centering
\vspace{-0.15in} {\scalebox{0.78}[0.78]{\includegraphics{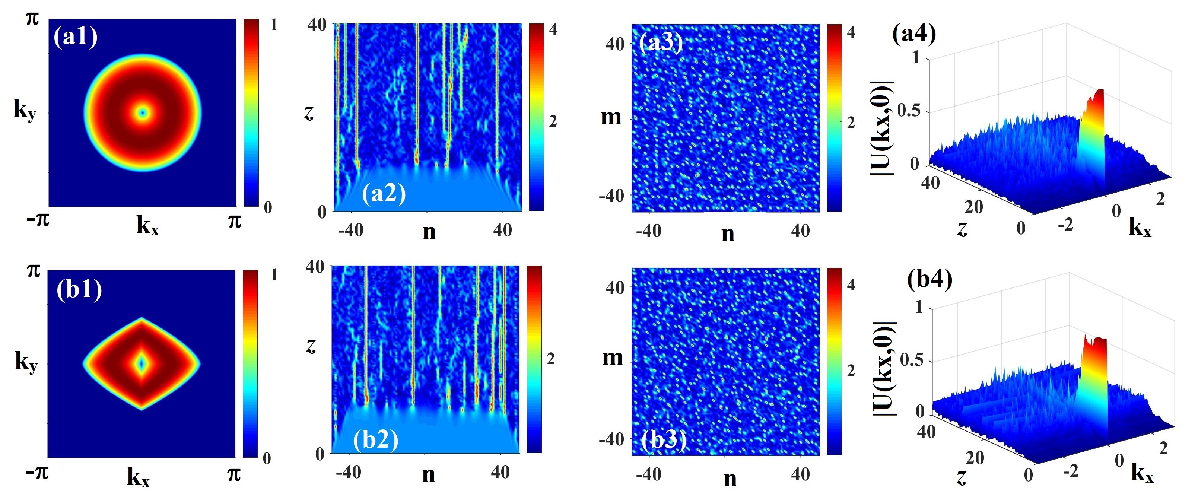}}}\hspace{-0.3in%
} \vspace{-0.1in}
\caption{MI produced by Eqs.~(\protect\ref{2D}) (top) and (\protect\ref%
{FNLSE2}) (bottom) with LIs $\protect\alpha =1,\protect\beta =1.5$ and the
linear-coupling coefficient $C=1$. (a1,b1) The instability growth rates $%
G_{1}$ and $G_{2}$ corresponding to Eq.~(\protect\ref{GR}). Simulations of
Eqs.~(\protect\ref{2D}) and (\protect\ref{FNLSE2}) with input $u_{n,m}(0)=1+%
\protect\varepsilon _{n,m}$: (a2,b2) The cross-section of amplitude $%
|u_{n,0}(z)|$, where $0\leq z\leq 40$. S (a3,b3): The final space-time
pattern at $z=40$. (a4,b4): Similar to (a2,b2), but in the Fourier space. }
\label{MIP2}
\end{figure}

Notably, these DRs align with that of the continuous fractional NLS
equation, except for the range of the wavenumber $\mathbf{k}$. Furthermore,
they differ from the DR of the standard 2D DNLS equation with the
nearest-neighbor coupling~\cite{KRB,Ke09,AC4},
\begin{equation}
i\frac{du_{n,m}}{dz}=C\left(
4u_{n,m}-u_{n+1,m}-u_{n-1,m}-u_{n,m+1}-u_{n,m-1}\right) -g\left\vert
u_{n,m}\right\vert ^{2}u_{n,m},  \label{DNLS}
\end{equation}%
which is
\begin{equation}
\omega _{3}(\mathbf{k})=2C[2-\cos (k_{x})-\cos (k_{y})].
\end{equation}%
Thus, the lattice band, defined by extremities of the DRs as specified by
Eqs.~(\ref{om}) and (\ref{om2}), spans the following ranges:
\begin{equation}
0\leq \omega _{1}(\mathbf{k})\leq C\left( \sqrt{2}\pi \right) ^{\alpha
},\qquad 0\leq \omega _{2}(\mathbf{k})\leq C\left( \pi ^{\alpha }+\pi
^{\beta }\right) .  \label{sp}
\end{equation}%
Discrete solitons can emerge in the form described by Eq.~(\ref{sp}) at
frequencies outside these bands. Shown in Fig.~\ref{LR} are the different
types of the linear DRs for $\omega _{j}(\mathbf{k})(j=1,2,3)$ with certain
parameters. The DRs for Eq.~\eqref{2D} are displayed in Fig.~\ref{LR}(a1) as
$\omega _{1}(k)$ with LI $\alpha =1$, and for Eq.~\eqref{FNLSE2} the DRs are
displayed in Fig.~\ref{LR}(a2) as $\omega _{2}(k)$ with LIs $\alpha
=2,\,\beta =1$. For Eq.~(\ref{DNLS}), the DRs are displayed in Fig.~\ref{LR}%
(a3) as $\omega _{3}(k)$, with the fixed coupling \ coefficient $C=1$. They
manifest different shapes: $\omega _{1}(\mathbf{k})$ features a cone, as per
Eq.~(\ref{oms}), while $\omega _{2}(\mathbf{k})$ corresponds to a parabolic
form in one direction and a straight line in the other one; lastly, $\omega
_{3}(\mathbf{k})$ exhibits a (cell of the) periodic structure.

\section{Modulation instability (MI) and excitation of rogue waves (RWs)}

Next, we address MI as the instability of CWs in the framework of the full
fractional DNLS equations (\ref{2D}) and (\ref{FNLSE2}). Our initial
approach considers the modulational perturbation of the spatially uniform
CW, $u_{n,m}(z)=e^{igz}$, as
\begin{equation}
u_{n,m}^{\epsilon }(z)=\left( 1+\varepsilon W_{n,m}(z)\right) e^{igz},
\label{pw}
\end{equation}%
with $\varepsilon \ll 1$. Substituting Eq.~\eqref{pw} in Eqs.~(\ref{2D}) and
(\ref{FNLSE2}), and linearizing with respect to $\varepsilon $, we derive
the following equations for perturbation amplitudes $W_{n,m}(z)$:
\begin{equation}
i\frac{dW_{n,m}}{dz}=C\left\{ \left( -\frac{\widehat{\partial }^{2}}{%
\widehat{\partial }n^{2}}-\frac{\widehat{\partial }^{2}}{\widehat{\partial }%
m^{2}}\right) ^{\alpha /2}W\right\} _{n,m}-g(W_{n,m}+W_{n,m}^{\ast })=0,
\label{MI1}
\end{equation}%
for Eq.~(\ref{2D}) and
\begin{equation}
i\frac{dW_{n,m}}{dz}=C\left\{ \left[ \left( -\frac{\widehat{\partial }^{2}}{%
\widehat{\partial }n^{2}}\right) ^{\alpha /2}+\left( -\frac{\widehat{%
\partial }^{2}}{\widehat{\partial }m^{2}}\right) ^{\beta /2}\right]
W\right\} _{n,m}-g(W_{n,m}+W_{n,m}^{\ast })=0,  \label{MI2}
\end{equation}
for Eq.~(\ref{FNLSE2}). We further assume that $W_{n,m}$ is represented by
the lowest Fourier modes as
\begin{eqnarray}  \label{Wmn}
W_{n,m}(z)=f_{+}e^{i(\mathbf{k}\cdot \mathbf{r}-\Omega(\mathbf{k})z)}
+f_{-}e^{-i(\mathbf{k}\cdot \mathbf{r}-\Omega(\mathbf{k}) z)}.
\end{eqnarray}
Incorporating ansatz \eqref{Wmn} in the linearized equations (\ref{MI1}) and
(\ref{MI2}), we obtain a relation between the perturbation frequency and
wavenumber, \textit{viz}.,

\begin{equation}
\Omega ^{2}(\mathbf{k})=C|\mathbf{k}|^{\alpha }\left( C|\mathbf{k}|^{\alpha
}-2g\right)  \label{MIS1}
\end{equation}%
for Eq.~(\ref{2D}) and
\begin{equation}
\Omega ^{2}(\mathbf{k})=C\left( |k_{x}|^{\alpha }+|k_{y}|^{\beta }\right) %
\left[ C\left( |k_{x}|^{\alpha }+|k_{y}|^{\beta }\right) -2g\right]
\label{MIS2}
\end{equation}%
for Eq.~(\ref{FNLSE2}).  The derivation of these relations is
given in Appendix B. Observing that the criterion for the emergence of MI
is $\Omega ^{2}<0$, we can deduce wavenumber conditions for the instability.
Note that Eq.~(\ref{MIS1}) is similar to the continuous 2D fractional NLS
equation, differing only in the wavenumber. As a result, the range of
unstable wavenumbers is the same, namely, $|\mathbf{k}|^{\alpha }<2g/C$. The
analysis naturally corroborates the absence of MI in the defocusing regime,
hence we address the self-focusing case, setting $g=1$. The wavenumber
condition for Eq.~(\ref{MIS2}) can be derived similarly. We thus identify
the instability growth rates $G_{1}$ and $G_{2}$ (the imaginary part of $%
\Omega $), which are associated with Eq.~(\ref{MIS1}) and Eq.~(\ref{MIS2}),
respectively:
\begin{equation}
\begin{array}{l}
G_{1}=\sqrt{C|\mathbf{k}|^{\alpha }\left( 2-C|\mathbf{k}|^{\alpha }\right) },%
\vspace{0.1in} \\
G_{2}=\sqrt{C\left( |k_{x}|^{\alpha }+|k_{y}|^{\beta }\right) \left[
2-C\left( |k_{x}|^{\alpha }+|k_{y}|^{\beta }\right) \right] }.%
\end{array}
\label{GR}
\end{equation}%
The expression for $G_{1}$ demonstrates that the largest instability growth
rate is $1$, achieved on a circle of radius $\left( 1/C\right) ^{\alpha }$
in the wavenumber space. The situation for $G_{2}$ is similar, except for
that the largest instability growth rate is attained at $|k_{x}|^{\alpha
}+|k_{y}|^{\beta }=1/C$.

\begin{figure}[!t]
\centering
\vspace{-0.15in} {\scalebox{0.7}[0.7]{\includegraphics{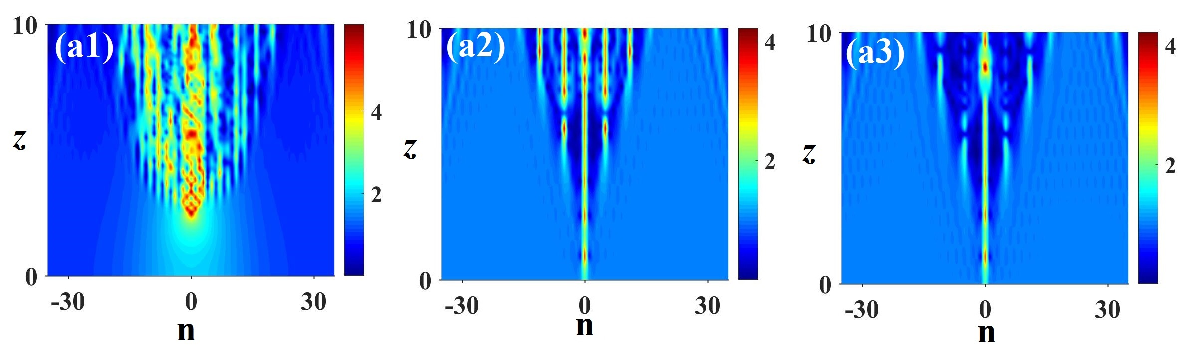}}}\hspace{-0.3in}
\vspace{-0.1in}
\caption{The RW evolution produced by simulations of Eq.~(\protect\ref{2D})
with parameters $\protect\alpha =1.5,\,C=1$ and input given by Eq.~(\protect
\ref{In1}), with different widths: (a1) $w=5$, (a2) $w=0.5$, (a3) $w=0.1$.
The evolution is displayed in cross section $m=0$.}
\label{RWG}
\end{figure}
\begin{figure}[!t]
\centering
\vspace{0.1in} {\scalebox{0.7}[0.7]{\includegraphics{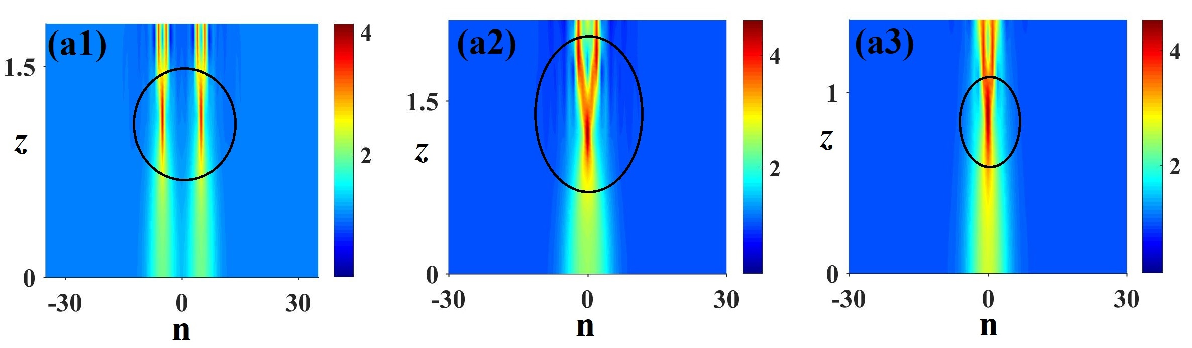}}}\hspace{-0.3in}
\vspace{-0.1in}
\caption{Different types of RWs produced by simulations of Eq.~(\protect\ref%
{2D}) with parameters $\protect\alpha =1.5,C=1$ and input~(\protect\ref{In2}%
), with $w=2.5$, for different center points: (a1) $n_{0}=5$, (a2) $n_{0}=2$%
, (a3) $n_{0}=1.5$. The evolution is displayed in cross section $m=0$. The
black elliptical regions delineate the high-amplitude RWs.}
\label{RWT}
\end{figure}
A quintessential example for the MI growth rate is produced in Figs. \ref%
{MIP2}(a1,b1), for $\alpha =1,\beta =1.5,C=1$. To corroborate the MI
prediction, we conduct numerical simulations using the initial condition
\begin{equation}
u_{n,m}(0)=1+\varepsilon _{n,m},\quad |\varepsilon _{n,m}|\ll 1.
\end{equation}%
It is observed that small perturbations feature exponential amplification,
until MI saturates due to the nonlinearity of the growing perturbation. The
amplification of the perturbation in the real space is accompanied by the
emergence of large-amplitude structures, as seen in Figs.~\ref{MIP2}(a2,a3),
where Fig.~\ref{MIP2}(a2) presents a cross-sectional view at $m=0$ and Fig.~%
\ref{MIP2}(a3) displays the spatiotemporal evolution up to $z=40$. Figures~%
\ref{MIP2}(b2,b3) exhibit similar outcomes. Conversely, in the Fourier
space, it is seen that sidebands around the original wavenumber are excited
(see Figs.~\ref{MIP2}(a4,b4)), resulting in the generation of harmonics.
Note that the amplitude is normalized in this analysis. A more detailed
examination of the relationship between MI and LI will be reported elsewhere.

To rigorously explore RW solutions on the fractional lattice, we perform
numerical investigation of Eqs.~(\ref{2D}) and (\ref{FNLSE2}). To this end,
we employ a Gaussian input, which is known as an effective method for
triggering RW phenomena through the gradient-catastrophe mechanism in the
focusing NLSE, as initially shown in the framework of the semiclassical
approximation for the continuum NLS equation~\cite{Be13}, and further
corroborated by experimental findings in nonlinear optics ~\cite{Ti17}.

%{\Huge [It is necessary to define what is shown by ovals in Fig. 4.]}

As an example, we do it for Eq.~(\ref{2D}) with two different inputs (the
results for Eq.~(\ref{FNLSE2}) are quite similar):

\begin{figure}[!t]
\centering
\vspace{-0.15in} {\scalebox{0.78}[0.78]{\includegraphics{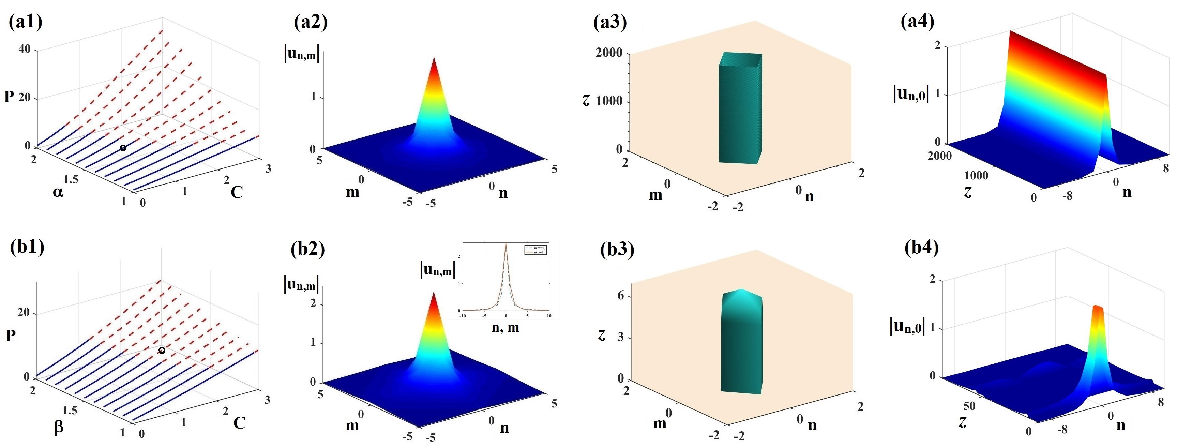}}}\hspace{%
-0.3in} \vspace{-0.1in}
\caption{Fundamental solitons produced by the isotropic equation (\protect
\ref{STA}) (top) and anisotropic one (\protect\ref{STA2}) (bottom),
respectively. (a1) Power $P$ vs. the coupling constant $C$ and LI $\protect%
\alpha $, as produced by the numerical solutions of Eq.~(\protect\ref{STA}),
with solid and dashed lines denoting stable and unstable segments,
respectively. (a2,a3): The profile and stable evolution of the soliton with $%
C=1$ and $\protect\alpha =1.5$. (a4): The evolution in cross-section $m=0$.
(b1) Power $P$ vs. the coupling constant $C$ and LI $\protect\beta $, as
produced by the numerical solutions of Eq.~(\protect\ref{STA2}) with fixed
LI, $\protect\alpha =1$. (b2,b3): The profile and unstable evolution of the
soliton with $C=2$ and $\protect\beta =1.5$. (b4): The evolution in
cross-section $m=0$. The inset in (b2) displays the cross-section
corresponding to $v(n,0)$ and $v(0,m)$. }
\label{FunS}
\end{figure}

\begin{itemize}
\item {} First, we consider Eq.~(\ref{2D}) with fixed parameters $\alpha
=1.5,C=1$, and the following initial condition:
\begin{equation}
u_{n,m}(0)=1+\exp \left( -\frac{n^{2}+m^{2}}{2w^{2}}\right) ,  \label{In1}
\end{equation}
where $w$ is the width of the input. The constant term $1$ represents the
normalized amplitude of the background CW, as the reference value.  The choice of this input, built as the Gaussian perturbation
added to the CW background, facilitates the evolution towards RWs. The
variation of the wave field for different values of $w$ are illustrated in
Fig.~\ref{RWG}, where the panels from left to right correspond to $w=5$, $%
w=0.5$, and $w=0.1$, respectively. These findings reaffirm the generality of
the gradient catastrophe scenario presented in Ref.~\cite{Be13}. Although
the nonintegrability of the lattice model distorts the resulting
Peregrine-soliton patterns, deviating from the \textquotedblleft Christmas
tree" structures produced in the paradigmatic NLS equation, this pattern
remains recognizable. As $w$ decreases, the RW pattern gradually evolves
towards a breathing-type solution. This observation is consistent with the
analysis for the integrable NLS equation outlined in Ref.~\cite{Be13}, which
has been similarly validated in various continuous~\cite{Ch18} and discrete~%
\cite{Ho10} systems. Similar results are produced by the anisotropic model
based on Eq.~(\ref{FNLSE2}) (not shown here).

\item {} Second, it is worthy to note is that we can generate higher-order
RWs from a superposition of multiple Gaussians. To this end, we numerically
solve Eq.~(\ref{2D}) with parameters $\alpha =1.5,C=1$, using the following
input including two Gaussians with width $w$ and centers placed at points $%
n=\pm n_{0}$ (including the input with virtual centers if $n_{0}$ is not
integer):
\begin{equation}
u_{n,m}(0)=1+\exp \left(-\frac{\left( n-n_{0}\right) ^{2}}{2w^{2}}\right)
+\exp \left(-\frac{\left( n+n_{0}\right) ^{2}}{2w^{2}}\right).  \label{In2}
\end{equation}
Here, we consider motion along the $n $-direction, although movement in the $%
(n, m)$ direction could also be explored. Different forms of RWs generated
by this input with $w=2.5$ are displayed in Fig.~\ref{RWT}. Initially, for $%
n_{0}=5$, a symmetric RW is obtained in Fig.~\ref{RWT}(a1). As $n_{0}$ is
reduced to $2$, a tri-RW state emerges, where the wave initially
concentrates into a single RW before splitting into two separate ones in
Fig.~\ref{RWT}(a2). Further decreasing $n_{0}$ to $1.5$ reveals a
significantly stronger tri-RW, as shown in Fig.~\ref{RWT}(a3).
%{\Huge [It was written here:
%\textquotedblleft }Note the contrast between Figs.~\ref{RWT}(a2) and (a3):
%the former one features a pronounced amplitude in the split components,
%whereas the latter lacks such structure.{\Huge " I see that both (a2) and
%(a3) feature \textquotedblleft }a pronounced amplitude in the split
%components{\Huge ". I suggest to delete this sentence, as it may easily
%provoke a question of reviewers.]}
\end{itemize}

\section{2D nonlinear modes and their stability}

In this section, we conduct a comprehensive investigation of fundamental and
vortical solitons produced by the isotropic and anisotropic models based on
Eqs.~(\ref{2D}) and (\ref{FNLSE2}). The analysis encompasses their existence
conditions and stability properties.

Localized stationary modes are looked for in the usual form,
\begin{eqnarray}
u_{n,m}(z)=v_{n,m}e^{-i\omega z},\quad v_{n,m}=v(n,m)
\end{eqnarray}
for the 2D fractional DNLS equations (\ref{2D}) and (\ref{FNLSE2}) with $g=1$
(the normalized self-focusing nonlinearity) and real frequency $\omega $.
Substituting this in Eqs.~(\ref{2D}) and (\ref{FNLSE2}) leads to the
stationary equations,
\begin{equation}
C\left( -\frac{\widehat{\partial }^{2}}{\widehat{\partial }n^{2}}-\frac{%
\widehat{\partial }^{2}}{\widehat{\partial }m^{2}}\right) ^{\alpha
/2}v_{n,m}-|v_{n,m}|^{2}v_{n,m}=\omega v_{n,m},  \label{STA}
\end{equation}%
and
\begin{equation}
C\left[ \left( -\frac{\widehat{\partial }^{2}}{\widehat{\partial }n^{2}}%
\right) ^{\alpha /2}+\left( -\frac{\widehat{\partial }^{2}}{\widehat{%
\partial }m^{2}}\right) ^{\beta /2}\right] v_{n,m}-|v_{n,m}|^{2}v_{n,m}=%
\omega v_{n,m}.  \label{STA2}
\end{equation}%
In these equations, $\omega $ can be scaled out by defining $v_{n,m}\equiv
\sqrt{-\omega }\hat{v}_{n,m}$ and $C\equiv -\omega \hat{C}$. Henceforth, we
fix $\omega =-1$ ($\omega <0$ is necessary for the creation of
bright-soliton solutions).

\begin{figure}[t]
\centering
\vspace{-0.05in} {\scalebox{0.72}[0.72]{\includegraphics{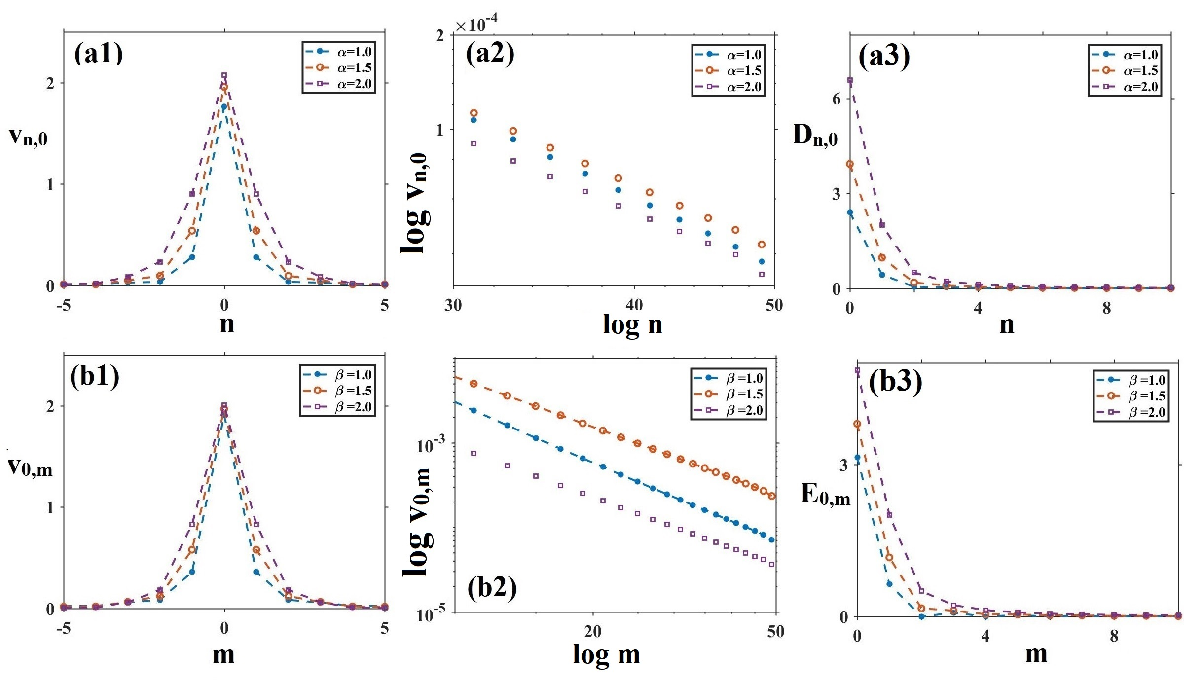}}}\hspace{-0.3in}
\vspace{-0.1in}
\caption{Solitons produced by Eq.~(\protect\ref{STA}) with fixed $C=1$, and
different typical values of LI: $\protect\alpha =1,1.5,2.0$: (a1)
Cross-section $v_{n,0}$; (a2) tails of the soliton profiles from panel (a1)
on the log-log scale; (a3) the effective coupling constant between lattice
sites $\left( 0,0\right) $ and $\left( n,0\right) $. According to the data
displayed in panel (a2), the decaying tail may be approximated by $%
v_{n,0}\sim |n|^{-\protect\eta }$ with $\protect\eta =1.80,1.77,1.67$
corresponding to $\protect\alpha =1,1.5,2.0$, respectively. Solitons
produced by (anisotropic) Eq.~(\protect\ref{STA2}) with fixed $\protect%
\alpha =1,\,C=1$, and typical values of the second LI: $\protect\beta %
=1,1.5,2.0$: (b1) a typical cross-section $v_{0,m}$; (b2) tails of the
soliton profiles from panel (b1) on the log-log scale, the respective decay
powers being $\protect\eta =2.38,2.16,1.99$ for $\protect\beta =1,1.5,2.0$,
respectively; (b3) the effective coupling constant between lattices site $%
\left( 0,0\right) $ and $\left( 0,m\right) $.}\label{COF}
\end{figure}
%{\LARGE [It is necessary to add labels to the vertical and horizontal axes
%in panels (a2) and (b2) of Fig. \ref{COF}. It is also necessary to indicate
%(in the caption to Fig. \ref{COF}) particular values of power }${\LARGE \eta
%}${\LARGE \ in the algebraically decaying asymptotic tail }$\left\vert
%{\LARGE u}_{{\LARGE m,n}}\right\vert {\LARGE \sim }\left( m^{2}+n^{2}\right)
%^{{\LARGE -\eta /2}}${\LARGE , which approximately correspond to the lines
%displayed in Figs. \ref{COF} (a2) and (b2)].}

The 2D nonlinear fractional-difference equation~(\ref{STA}) can be solved
numerically, starting from the anti-continuum (ac) limit, $C=0$ \cite%
{AC3,AC4}. In this limit, the solution of Eq. (\ref{STA}) is obvious:
\begin{equation}
v_{n,m}^{(0)}=\left\{
\begin{array}{l}
e^{i\theta _{n,m}},\quad (n,m)\in S,\vspace{0.1in} \\
0,\quad (n,m)\in \mathbb{Z}^{2}\backslash S,%
\end{array}%
\right.  \label{ACS}
\end{equation}%
where $S$ is a finite set of nodes on the square lattice $\left( n,m\right)
\in \mathbb{Z}^{2}$ and $\theta _{n,m}$ are phases of the populated
(nonzero) sites. In particular, discrete solitons and vortices correspond to
$\theta _{n,m}=0$ and $\theta _{n,m}\in \left[ 0,2\pi \right] $,
respectively. The existence of the continuous family of solitons
branching from Eq.~(\ref{ACS}) at finite values of $C$ can be rigorously
proven using the implicit function theorem \cite{AC3,AC4}.

The linear stability analysis of the standing-wave solutions can be
performed in the framework of the linearized Bogoliubov-de Gennes equations
for small perturbations. Substituting the perturbed solution
\begin{equation}
u_{n,m}(z)=\left[ v_{n,m}+\epsilon \left( a_{n,m}e^{\lambda z}+b_{n,m}^{\ast
}e^{\lambda ^{\ast }z}\right) \right] e^{-i\omega z}
\end{equation}%
with $\varepsilon \ll 1$ in Eq. (\ref{2D}), one derives the eigenvalue
problem for the perturbation amplitudes $\left( a_{n,m},b_{n,m}\right) $ and
instability growth rate $\lambda $,
\begin{equation}
\left(
\begin{matrix}
L_{11} & L_{12}\vspace{0.1in} \\
-L_{12}^{\ast } & -L_{11}^{\ast }%
\end{matrix}%
\right) \!\!\left(
\begin{matrix}
a_{n,m}\vspace{0.1in} \\
b_{n,m}%
\end{matrix}%
\right) \!=\!i\lambda \left(
\begin{matrix}
a_{n,m}\vspace{0.1in} \\
b_{n,m}%
\end{matrix}%
\right) ,  \label{spectral}
\end{equation}%
where
\begin{equation}
\begin{array}{rl}
L_{11} & =\displaystyle C\left( -\frac{\widehat{\partial }^{2}}{\widehat{%
\partial }n^{2}}-\frac{\widehat{\partial }^{2}}{\widehat{\partial }m^{2}}%
\right) ^{\alpha /2}-2|v_{n,m}|^{2}-\omega ,\vspace{0.1in} \\
L_{12} & =-v_{n,m}^{2}.%
\end{array}
\label{op}
\end{equation}%
Similarly, we deduce the corresponding eigenvalue problem for Eq.~%
\eqref{FNLSE2}, with the different operator
\begin{equation}
L_{11}=C\left[ \left( -\frac{\widehat{\partial }^{2}}{\widehat{\partial }%
n^{2}}\right) ^{\alpha /2}-\left( \frac{\widehat{\partial }^{2}}{\widehat{%
\partial }m^{2}}\right) ^{\beta /2}\right] -2|v_{n,m}|^{2}-\omega .
\end{equation}%
The discrete soliton $u_{n,m}(z)$ is unstable if there exists at least a
single eigenvalue with Re$(\lambda )>0$. Predictions of the linear-stability
analysis are then verified by direct simulations of the perturbed evolution,
running them by dint of the fourth-order Runge-Kutta method~\cite{At91}. In
the course of the simulations, the conservation of Hamiltonian (\ref{Ha})
and power (\ref{Po}) was monitored to control the accuracy of the numerical
scheme.

%\subsection{Varational }
%\begin{equation}
%H_{1}=\frac{C}{2}\sum_{m,n}v_{n,m}\left( -\frac{\widehat{\partial }^{2}}{\widehat{\partial }n^{2}}
%-\frac{\widehat{\partial }^{2}}{\widehat{\partial}m^{2}}\right) ^{\alpha /2}v_{n,m}-\frac{%
%g }{4}\sum_{m,n}v_{n,m}^{4}-\frac{\omega}{2}\sum_{m,n}v_{n,m}^{2},  \label{La}
%\end{equation}%
%
%\bee
%\sum_{l_x}e^{i(l_x+n)k_x}e^{-\frac{(n+l_x)^2}{2W^2}}
%\ene
%
%\bee
%\sum_{n}e^{-ink_x}e^{-\frac{n^2}{2W^2}}
%\ene

\begin{figure}[!t]
\centering
\vspace{-0.05in} {\scalebox{0.75}[0.75]{\includegraphics{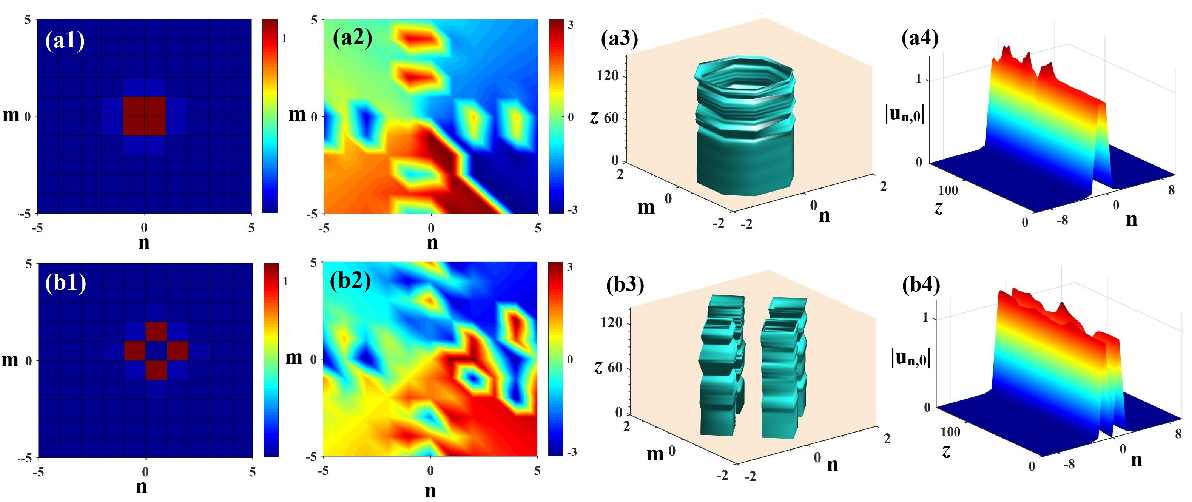}}}\hspace{-0.3in}
\vspace{-0.1in}
\caption{A typical unstable vortex soliton produced by Eq.~(\protect\ref{STA}%
) with LI $\protect\alpha =1$ and $C=0.1$. (a1,a2) The intensity and phase
patterns of an intersite-centered soliton. The unstable evolution is
displayed in panels (a3) and (a4). The second row is similar to the first
one, representing a typical unstable onsite-centered vortex soliton. }
\label{IU}
\end{figure}

\subsection{ Fundamental (zero-vorticity) solitons and their stability}

\begin{figure}[!t]
\centering
\vspace{-0.15in} {\scalebox{0.62}[0.62]{\includegraphics{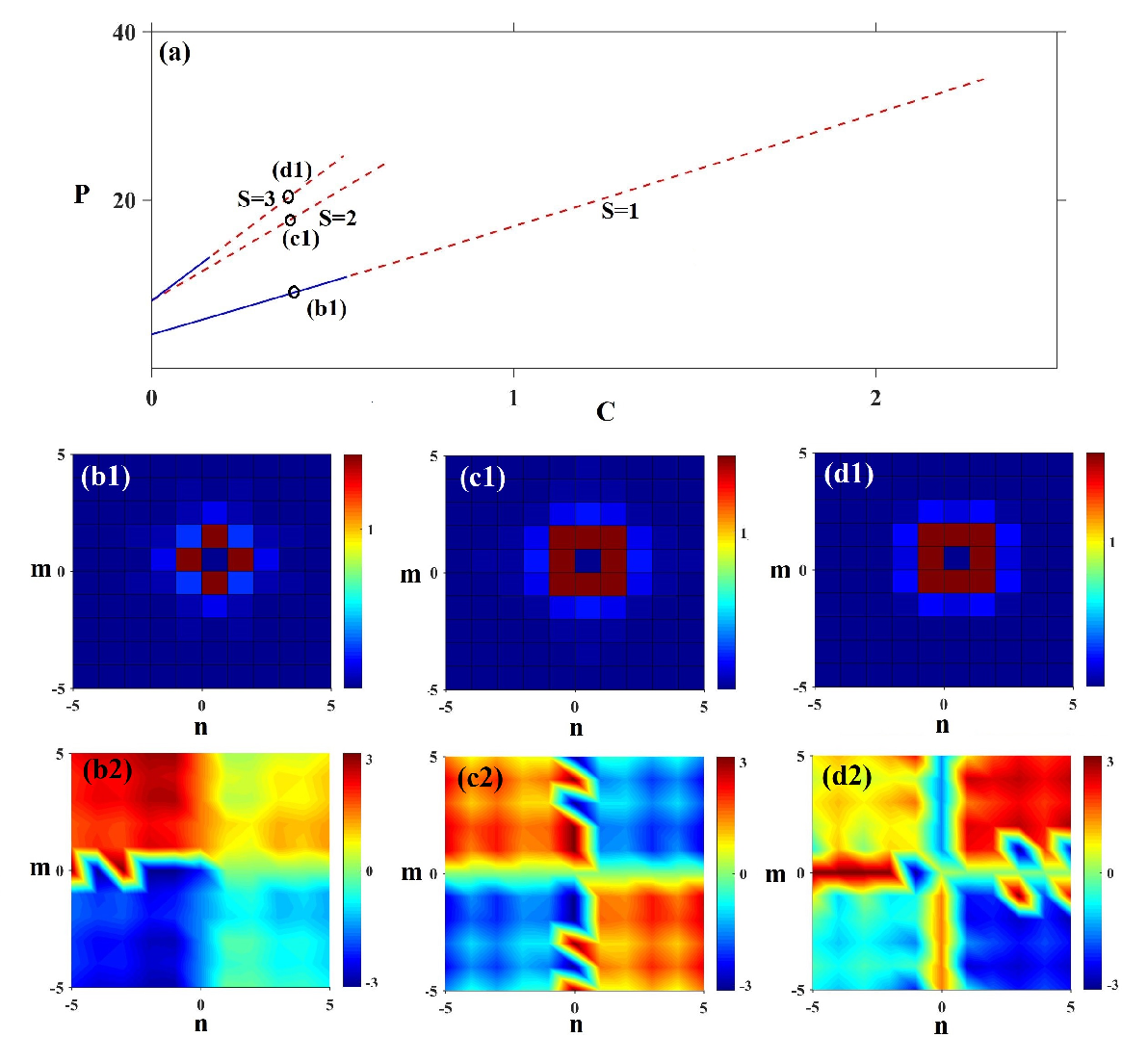}}}\hspace{-0.3in}
\vspace{-0.1in}
\caption{Quasi-isotropic vortex solitons produced by the symmetric equation (%
\protect\ref{STA2}) with LIs $\protect\alpha =\protect\beta =1$. (a) Power $%
P $ vs. $C$ with different vorticities $S=1,2,3$, where solid and dashed
lines denote stable and unstable families of the vortex solitons,
respectively. (b1,b2) The intensity and phase structures of the discrete
vortex soliton with $S=1$ at $C=0.4$. (c1,c2) The same as in (b1,b2) but for
$S=2$. (d1,d2) The same as in (b1,b2), but for $S=3$. }
\label{IV}
\end{figure}

\begin{figure}[!t]
\centering
\vspace{-0.02in} {\scalebox{0.7}[0.7]{\includegraphics{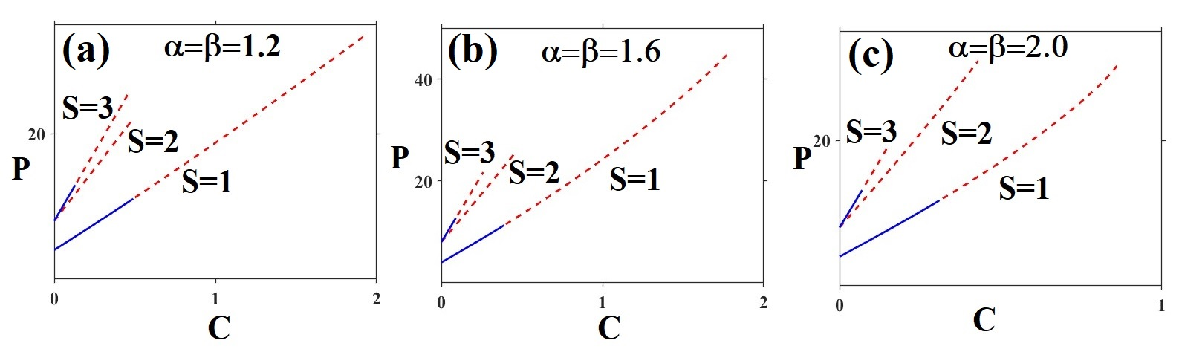}}}\hspace{-0.3in}
\vspace{-0.1in}
\caption{Quasi-isotropic vortex solitons produced by the symmetric equation (%
\protect\ref{STA2}) with different LIs at (a) $\protect\alpha =\protect\beta %
=1.2$, (b) $\protect\alpha =\protect\beta =1.6$ as well as at (c) $\protect%
\alpha =\protect\beta =2.0$. The blue solid line and the red dashed line
represent stable and unstable vortex solitons, respectively.}
\label{DQV}
\end{figure}

We begin the analysis with the existence and stability of fundamental
solitons produced by Eqs.~(\ref{STA}) and (\ref{STA2}), starting from the
anti-continuum (AC) limit, $C=0$~\cite{Ke09,AC}.

\vspace{0.1in} \textit{Case 1:\thinspace\ The isotropic model.}---Results
for fundamental isotropic solitons produced by Eq.~(\ref{STA}) are
summarized in Fig.~\ref{FunS}(a1) in the form of dependences of the total
power $P$ (see Eq. (\ref{Po})) on LI, $\alpha \in (1,2]$, and coupling
constant $C\in \lbrack 0,3]$. As mentioned above, in the continuous 2D
fractional NLS equation, solitons (including ones of the fundamental and
vortex types) are destabilized by the critical and supercritical collapse at
$\alpha =2$ and $\alpha <2$, respectively. However, the destabilization is
forestalled in the discrete setting, as shown in Fig.~\ref{FunS}(a1), where
the stability area is indicated by solid curves. It is seen that the
stability domain of the fundamental solitons increases with the decrease of
LI $\alpha $, which is explained by the weaker interactions between sites of
the discrete medium. To illustrate this feature in detail, we display the
effective coupling constant $D_{l_{x},l_{y}}^{\left( \alpha \right) }$
between sites $\left( 0,0\right) $ and $\left( n,0\right) $ with fixed $C=1$%
, for typical values of LI, $\alpha =1,1.5,2.0$, in Fig.~\ref%
{COF}(a3). It is seen that coefficient $D_{l_{x},l_{y}}^{\left( \alpha
\right) }$ indeed decreases with the decrease of LI $\alpha $. Another
conclusion suggested by Fig. \ref{FunS}(a1) is that the $P(C)$ curves shift
down as $\alpha $ decreases.

It is known that fractional solitons produced by the continuous fractional
NLS equations display a power-law decay~of their tails \cite{Kl14,Fr10,Guo18}.
As an illustrative example, one can take the 1D linear fractional
equation (\ref{FSE}) with $V(x)=0$ and substitute a trial Lorentzian profile
\begin{equation}
u_{\mathrm{Lorentz}}(x)=\left(x^{2}+x_{0}^{2}\right) ^{-1}e^{-i\mu t}
\end{equation}
in the fractional-diffraction\ term. Then, a
straightforward calculation of that term yields the following asymptotic
result, at $x^{2}\rightarrow \infty $:%
\begin{equation}
\left( \!-\frac{\partial ^{2}}{\partial x^{2}}\!\right) ^{\alpha /2}\!u_{%
\mathrm{Lorentz}}(x)\big|_{\mathrm{at~}x^{2}\gg x_{0}^{2}}\approx -\frac{\Gamma
\left( \alpha +1\right) }{8x_{0}}|x|^{-\left( \alpha +1\right) }e^{-i\mu t}
\end{equation}%
(here $x_{0}>0$ is adopted by definition). Taking into regard the first term
in Fig. (\ref{FSE}), one concludes that Eq. (\ref{FSE}) with $\alpha =1$
yields a self-consistent power-law tail $\sim |x|^{-2}$, the respective
chemical potential being $\mu =-\left( 8x_{0}\right) ^{-1}$. Note that the
numerically found decay rate of the tail in the 2D discrete system with $\alpha=1$,
$\eta=1.80$ (see the caption to Fig. \ref{COF}), is indeed close to value $\eta=2$ produced by
the present analysis for the 1D continuous fractional system.

For the present discrete fractional model, representative shapes of the solitons with
fixed $C=1$ are potted in Fig.~\ref{COF}(a1),  Their tails are
displayed in Fig.~\ref{COF}(a2) on the log-log scale. The results
demonstrate a power-law decay of their tails, with the decay rates (their
values are given in the caption to Fig. \ref{COF}) increasing with the
decrease of LI $\alpha $, which is again explained by the weaker coupling
between sites of the lattice, corresponding to smaller values of
coefficients $D_{l_{x},l_{y}}^{\left( \alpha \right) }$ (see Fig.~\ref{COF}%
(a3)). A typical profile of a stable 2D fundamental soliton with $\alpha
=1.5$ and $C=1$ is displayed in Fig.~\ref{FunS}(a2). Its stability is
corroborated by simulating its perturbed evolution, which is displayed in
Figs.~\ref{FunS}(a3,a4).

\vspace{0.1in} \textit{Case 2:\thinspace \thinspace\ The anisotropic model.}%
---To demonstrate results for fundamental solitons produced by the
anisotropic model based on Eq.~(\ref{STA2}), we focus on the characteristic
case with fixed LI $\alpha =1$ and varying the other LI, $\beta \in (1,2]$.
The dependence of power $P$ on\ $\beta $ and coupling constant $C$ is
presented in Fig.~\ref{FunS}(b1). Similar to the case of the isotropic model
(cf. Fig. \ref{FunS}(a1)), the stability region increases as $\beta $
decreases, as a consequence of the change of effective coupling constant $%
E_{lx,ly}^{\left( \alpha \right) }$ with fixed $C=1$, as shown
in Fig.~\ref{COF}(b3). The cross-section profiles of the fundamental
solitons are displayed in Fig.~\ref{COF}(b1) for different values of the
second LI $\beta $, and their tails are shown, on the log-log scale, in Fig.~%
\ref{COF}(b2), verifying the power-law decay of the tails.

A typical example of an unstable fundamental soliton in the anisotropic
model, with LIs $\alpha =1$,\thinspace\ $\beta =1.5$ and coupling constant $%
C=2$, is displayed in Fig.~\ref{FunS}(b2). Its unstable evolution is shown
in Figs.~\ref{FunS}(b3,b4), with the intensity decaying to zero.

\begin{figure}[!t]
\centering
\vspace{-0.15in} {\scalebox{0.5}[0.5]{\includegraphics{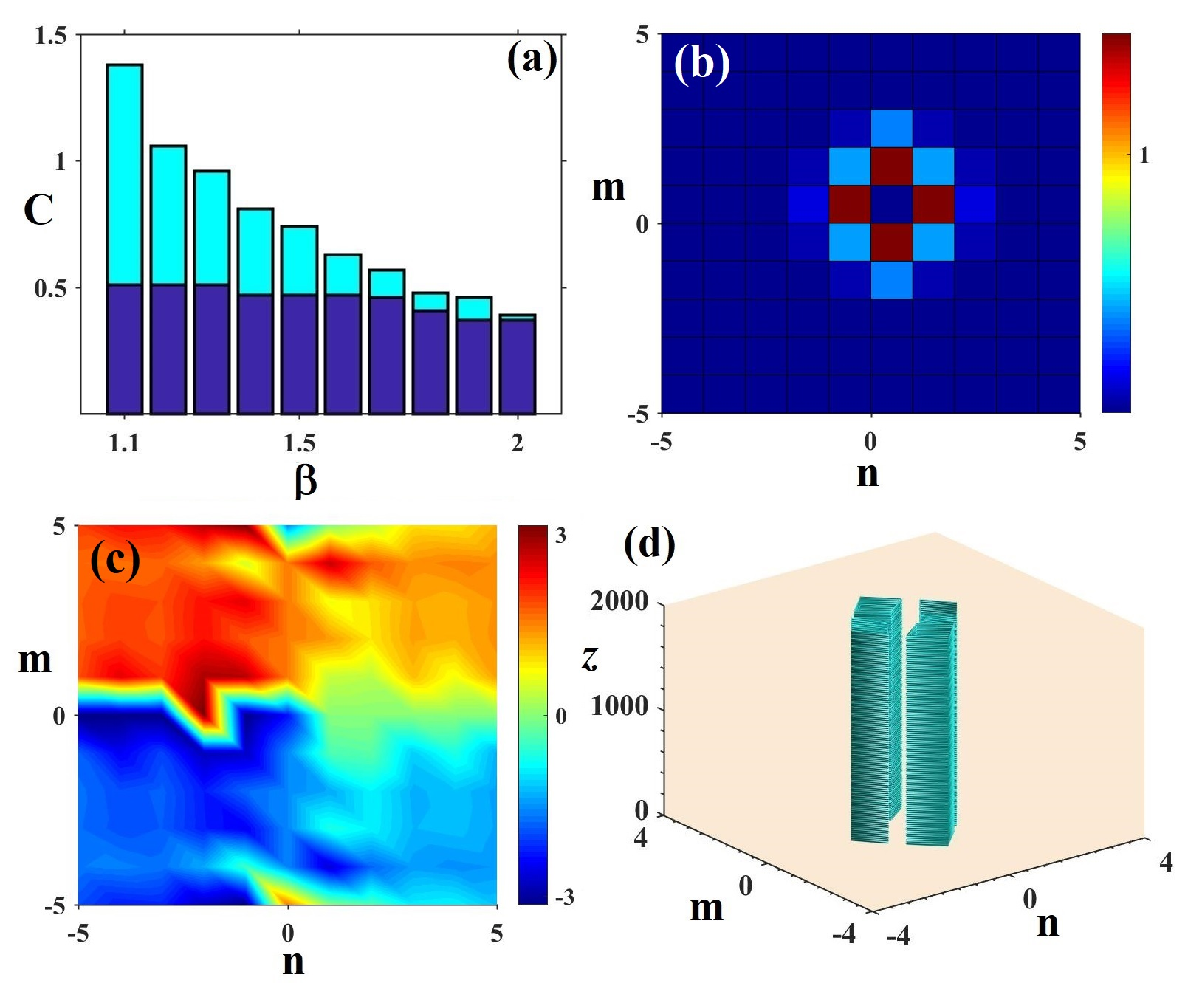}}}\hspace{-0.3in}
\vspace{-0.05in}
\caption{Anisotropic vortex solitons produced by the 2D asymmetric
fractional DNLS equation~(\protect\ref{STA2}) with $\protect\alpha \neq
\protect\beta $. (a) The existence and stability regions of the discrete
vortex solitons with $S=1$, for different values of LI $\protect\beta $ and
fixed $\protect\alpha =1$. Green and blue bars represent the existence and
stability regions, respectively. (b,c) The intensity and phase structure of
the stable discrete vortex soliton with $S=1$, $C=0.3$ {\color{red} as well
as $\protect\beta=2$}. (d) The stable perturbed evolution of the same vortex
soliton. }
\label{AIV}
\end{figure}

\subsection{Vortex solitons and their stability}

The consideration of vortex solitons with integer winding numbers $S=1,2,3,$
..., is another natural subject in the study of the 2D fractional DNLS
models.

First, extensive simulations of the isotropic equation~(\ref{STA})
demonstrate that this model readily supports stationary vortex solitons, but
they are all unstable. Typical examples of unstable intersite- and
onsite-centered vortex solitons with $S=1$ are displayed in Fig.~\ref{IU}
for $\alpha =1$ and $C=0.1$.  As usual, the phase singularity of
a vortex occurs at the point where the amplitude of the solution is
vanishing. Simulations of the perturbed evolution of these solitons,
displayed in Figs.~\ref{IU}(a3,a4) and (b3,b4), demonstrate their
spontaneous transformation into breathers  that perform
apparently randomized oscillations. This is different from the usual
behavior of unstable vortex solitons, which demonstrate splitting into
separating fragments \cite{Boris01,AC4}.

Next, the stability of quasi-isotropic vortex solitons of the symmetric
equation~(\ref{STA2}) with equal LIs, $\alpha =\beta =1$, and different
vorticities $S=1,2,3$ are summarized in Fig.~\ref{IV}(a). Naturally, the
region for the coupling constant $C$ supporting the existence of the vortex
solitons shrinks with the increase of $S$.  The curves in Fig. %
\ref{IV}(a) abort when they hit the boundary of the existence area of the
vortex solitons. We find \emph{stable} quasi-isotropic vortex solitons with
winding numbers $S=1$ and $3$, while ones with $S=2$ are completely
unstable, similar to what has been reported in previously studied
non-fractional lattice models with the nearest-neighbor coupling \cite%
{Boris01,AC4}. Some representative examples of the local-power and phase
structure of the vortex solitons are shown for $C=0.4$ in Figs.~\ref{IV}%
(b1,b2,c1,c2,d1,d2). The comparison with previous (non-fractional) models~%
\cite{Boris01,AC3} suggests that the present fractional one, based on Eq.~(%
\ref{STA2}), gives rise to a broader stability area for the vortex solitons.
Additionally, we have considered quasi-isotropic vortex solitons
in the model with different LIs, and the topological charge ranging from $1$
to $3$, see the results in Fig.~\ref{DQV}. One observes in the figure that
both the existence and stability ranges of the solitons decrease as $\alpha $
increases, similar to the above-mentioned results for the fundamental
solitons, primarily due to the change in the effective coupling coefficient.

Finally, we consider the asymmetric vortex solitons produced by the
anisotropic equation~(\ref{STA2}), with different LIs $\alpha \neq \beta $.
Results are summarized in Fig.~\ref{AIV}(a) for different values of LI $%
\beta $, fixing $\alpha =1$. Only discrete vortex solitons with $S=1$ are
found in this model. Their existence and stability regions, represented by
green and blue bars, respectively, in Fig.~\ref{AIV}(a), shrink as $\beta $
decreases. A typical stable onsite-centered vortex soliton with $S=1$ is
presented for $C=0.3$ in Figs.~\ref{AIV}(b,c), where the soliton's
anisotropy is clearly seen. The stable perturbed evolution of the same
vortex soliton is demonstrated in Fig.~\ref{AIV}(d) by means of isosurfaces.

\section{Conclusions and discussions}

In the present work, we have proposed the novel models of 2D fractional
dynamical lattices, based on fractional DNLS equations. The new discrete
versions of the quasi-RFD (Riesz fractional derivative) and Laplacian,
characterized by the respective LIs (L\'{e}vy indices), are naturally
introduced by means of the respective direct and inverse Fourier transform.
One 2D fractional DNLS equation includes the isotropic fractional Laplacian
with LI $\alpha \in (1,2]$. The anisotropic model makes use of the
fractional DNLS equation with the discrete derivatives acting independently
on two spatial coordinates in the 2D lattice, each one characterized by its
own LI, $\alpha $ and $\beta $, and respective lattice coupling constants, $%
C_{\alpha }$ and $C_{\beta }$. The symmetric version of the latter model,
with $\alpha =\beta $ and $C_{\alpha }=C_{\beta }$, is considered too. The
non-fractional version of the models, with LI $=2$, introduces the novel
lattice system, which includes the long-range couplings $\sim (-1)^{l}l^{-2}$
in the two directions between lattice sites separated by integer distance $l$%
. Utilizing this definition, the exact linear DRs (dispersion relations) and
MI (modulational instability) of CWs (continuous waves, alias plane waves)
are rigorously derived, exhibiting congruence with their continuous analogs.
The generation of RWs (rogue waves) has been investigated by means of
simulations of the underlying equations with Gaussian inputs.

Unlike the continuum fractional NLS equations in the 2D space, the
underlying lattice structure arrests the onset of the collapse, thus making
it possible to predict stable fundamental and vortical solitons. The
formation and stability of the soliton families are explored in detail,
starting from the ac (anti-continuum) limit. The stability is established by
means of the linear-stability analysis and verified by direct simulations.
On the contrary to the situation in the continuum limit, when smaller LI
makes the setting more prone to the onset of the collapse and resulting
destabilization of solitons, in the discrete systems smaller LI values favor
the stability of the solitons, which is explained by weaker coupling between
the lattice sites. The isotropic model supports stable fundamental solitons,
while all the vortical modes are unstable. On the other hand, the symmetric
system with independent fractional derivatives acting along the two discrete
coordinates maintains stable vortex-soliton families with winding numbers $%
S=1$ and $3$. Furthermore, the anisotropic system with $\alpha \neq \beta $
produces stable vortex solitons with $S=1$.

As an extension of the analysis, it may be interesting to
consider the 2D discrete system on more sophisticated underlying lattices,
such as triangular, hexagonal, and quasiperiodic ones. Similarly, it may be
relevant to consider the 1D discrete system based on a quasiperiodic 1D
lattice.

Another relevant direction may be the investigation of 2D fractional media
with the semi-discrete structure~\cite{Zh19}, i.e., the continuum fractional
derivative acting along one coordinate, and the discrete fractional
derivative acting in the perpendicular direction, the respective LIs being
different too. For example, the 2D fractional semi-discrete NLS equation is
\begin{equation}
i\frac{du_{m}}{dz}=\left[ C_{\alpha }\left( -\frac{\partial ^{2}}{\partial
x^{2}}\right) ^{\alpha /2}+C_{\beta }\left( -\frac{\widehat{\partial }^{2}}{%
\widehat{\partial }m^{2}}\right) ^{\beta /2}\right]
u_{m}+V(x,m)u_{m}+F(x,m,|u_{m}|^{2})u_{m}\text{,}  \label{FNLSE2g}
\end{equation}%
where $u_{m}=u(x,m,z)$ is an envelope field of continuous variables $%
x,\,z\in \mathbb{R}$ and discrete one $m\in \mathbb{Z}$, the LIs $\alpha
,\,\beta \in (1,2]$,\thinspace\ $F(\cdot )$ is a function of $%
x,m,|u_{m}|^{2} $, and $V(x,m)$ is a real or complex ($\mathcal{PT}$%
-symmetric) external potential.

A challenging possibility is to implement the fractional discrete setting in
the 3D geometry -- for example, in the form of the 3D isotropic fractional
DNLS equation,
\begin{equation}
i\frac{du_{n,m,s}}{dz}=C\left\{ \left( -\frac{\widehat{\partial }^{2}}{%
\widehat{\partial }n^{2}}-\frac{\widehat{\partial }^{2}}{\widehat{\partial }%
m^{2}}-\frac{\widehat{\partial }^{2}}{\widehat{\partial }s^{2}}\right)
^{\alpha /2}u\right\}
_{n,m,s}+V(m,n,s)u_{n,m,s}+F(n,m,s,|u_{n,m,s}|^{2})u_{n,m,s},  \label{3D}
\end{equation}%
and the 3D anisotropic fractional DNLS equation,
\begin{equation}
\begin{array}{rl}
i\dfrac{du_{n,m,s}}{dz}= & \displaystyle\left\{ \left[ C_{\alpha }\left( -%
\frac{\widehat{\partial }^{2}}{\widehat{\partial }n^{2}}\right) ^{\alpha
/2}+C_{\beta }\left( -\frac{\widehat{\partial }^{2}}{\widehat{\partial }m^{2}%
}\right) ^{\beta /2}+C_{\gamma }\left( -\frac{\widehat{\partial }^{2}}{%
\widehat{\partial }s^{2}}\right) ^{\gamma /2}\right] u\right\} _{n,m,s}%
\vspace{0.1in} \\
& \quad +V(m,n,s)u_{n,m,s}+F(n,m,s,|u_{n,m,s}|^{2})u_{n,m,s}\text{,}%
\end{array}
\label{FNLSE3}
\end{equation}%
where $u_{n,m,s}=u(n,m,s,z)$ is an envelope field of continuous variable $%
z\in \mathbb{R}$ and discrete ones $n,m,s\in \mathbb{Z}$, the LIs are $%
\alpha ,\,\beta,\, \gamma \in (1,2]$,\thinspace\ $F(\cdot )$ is a function
of $n,m,s,|u_{n,m,s}|^{2}$, and $V(n,m,s)$ is a real or complex external
potential. \newline

\vspace{0.1in} \noindent\textbf{ACKNOWLEDGMENTS} %\vspace{0.1in}

%\acknowledgments

The work of Z.Y. was supported by the National Natural Science Foundation of
China (Nos. 11925108 and 12471242). The work of B.A.M. is supported, in part, by grant No.
1695/22 from the Israel Science Foundation.

\vspace{0.1in}\vspace{0.1in} \noindent\textbf{CONFLICT OF INTEREST STATEMENT}
%\vspace{0.1in}

We declare we have no competing interests.

\vspace{0.1in}\vspace{0.1in} \noindent\textbf{DATA AVAILABILITY STATEMENT}
%\vspace{0.1in}

The data that support the findings of this study are available on a
reasonable request from the corresponding author.

 %\appendix
\setcounter{equation}{0}
\renewcommand{\theequation}{A.\arabic{equation}}

\v\noindent{\bf Appendix A.\, Derivations of Remark 3 and Eq.~(\ref{oms}) from the main text:}

%\textit{\ Proof of Remark 3:}

\v For the first equality in Eq.~(\ref{proof}), we note that the
coupling coefficient $D_{l_{x},l_{y}}^{(\alpha )}$ in Eq.~(\ref{C1}) can be
written as
\begin{equation}
D_{l_{x},l_{y}}^{(\alpha )}=\frac{1}{4\pi ^{2}}\int_{-\pi }^{+\pi
}\int_{-\pi }^{+\pi }e^{i\left( k_{x}l_{x}+k_{y}l_{y}\right) }\left(
k_{x}^{2}+k_{y}^{2}\right) ^{\alpha /2}dk_{x}dk_{y},
\end{equation}%
\label{C3} due to the evenness/oddness of the trigonometric functions.
Considering the definition of the fractional derivatives in Eq.~(\ref{2dd}),
one can obtain
\begin{equation}
\begin{array}{rl}
\left\{ \left( -\dfrac{\widehat{\partial }^{2}}{\widehat{\partial }n^{2}}-%
\dfrac{\widehat{\partial }^{2}}{\widehat{\partial }m^{2}}\right) ^{\alpha
/2}e^{i\mathbf{k}\cdot \mathbf{r}}\right\} _{n,m} & =\displaystyle%
\sum_{l_{x},l_{y}=-\infty }^{+\infty }D_{l_{x},l_{y}}^{(\alpha
)}e^{i\left( k_{x}(n+l_{x})+k_{y}(m+l_{y})\right) } \\
& =\displaystyle\sum_{l_{x},l_{y}=-\infty }^{+\infty
}D_{l_{x},l_{y}}^{(\alpha )}e^{i\left( k_{x}l_{x}+k_{y}+l_{y}\right) }e^{i%
\mathbf{k}\cdot \mathbf{r}} \\
& =\left(
k_{x}^{2}+k_{y}^{2}\right) ^{\alpha /2}e^{i\mathbf{k}\cdot \mathbf{r}},%
\end{array}%
\end{equation}%
where the last equality comes from the symmetry of $\left(
k_{x}^{2}+k_{y}^{2}\right) ^{\alpha /2}(=|\mathbf{k}|^{\alpha })$ and periodicity. Similarly, the second
equality in Eq.~(\ref{proof}) can be also derived.

Eqs.~(\ref{om}) and (\ref{om2}) respectively yield the first and second expressions in Eq.~(\ref{oms}), closely aligning with the derivation  of Remark 3 mentioned above. For example, the second equality  in Eq.~(\ref{oms}) is derived from Eq. (\ref{om2}) by noting the expression for \( E_{l_{x}}^{(\alpha)} \) in Eq.~(\ref{C2}), i.e.,
\begin{equation}
E_{l_{x}}^{(\alpha )}= \dfrac{1}{\pi }\displaystyle\int_{0}^{\pi }\cos \left( k_{x}l_{x}\right)
k_{x}^{\alpha }dk_{x},
\end{equation}
as well as the expression for \( E_{l_{y}}^{(\beta)} \),
\begin{equation}
E_{l_{y}}^{(\beta )}= \dfrac{1}{\pi }\displaystyle\int_{0}^{\pi }\cos \left( k_{y}l_{y}\right)
k_{y}^{\beta }dk_{y}.
\end{equation}
Substituting them into Eq.~(\ref{om2}), and considering the symmetry and periodicity of \( |k_x|^{\alpha} \) and \( |k_y|^{\beta} \) over the interval \( [-\pi, \pi] \), we obtain
\begin{equation}
\omega _{2}(\mathbf{k})=C(|k_{x}|^{\alpha }+|k_{y}|^{\beta }).
\end{equation}
In fact, \( E_{l_{x}}^{(\alpha )} \) and \( E_{l_{y}}^{(\beta )} \) are the Fourier expansion coefficients of \( |k_x|^{\alpha} \) and \( |k_y|^{\beta} \) over the interval \( [-\pi, \pi] \), respectively. The first  equality  in Eq.~(\ref{oms}) can be derived from Eq. (\ref{om}) similarly.

\v\noindent {\bf Appendix B.\, Derivations of Eqs.~(\ref{MIS1}) and (\ref{MIS2}):}
\setcounter{equation}{0}
\renewcommand{\theequation}{B.\arabic{equation}}

\v Here we show how to derive Eq.~(\ref{MIS1}), the derivation of Eq.~(\ref%
{MIS2}) being similar. Substituting Eq.~(\ref{Wmn}) into Eq.~(\ref{MI1}) and
considering coefficients of $e^{i(\mathbf{k}\cdot \mathbf{r}-\Omega (\mathbf{%
k})z)}$ and $e^{-i(\mathbf{k}\cdot \mathbf{r}-\Omega (\mathbf{k})z)}$
separately, we obtain
\begin{equation}
\begin{array}{rl}
& \Omega f_{1}-C|\mathbf{k}|^{\alpha }f_{1}+gf_{1}+gf_{2}=0, \\[1em]
& \Omega f_{2}+C|\mathbf{k}|^{\alpha }f_{2}-gf_{2}-gf_{1}=0.%
\end{array}%
\end{equation}%
The derivation here also relies on the equation in Remark 3, for which we
provided a proof in the above text. The necessary and sufficient condition
for the above-mentioned homogeneous linear system to have a solution in
terms of $(f_{1},f_{2})$ is that the vanishing of the determinant of the
coefficient matrix. Thus we derive
\begin{equation}
\Omega ^{2}=(C|\mathbf{k}|^{\alpha }-1)^{2}-1,
\end{equation}%
which is precisely Eq.~(\ref{MIS1}).
%\section{Discretization of the coupling coefficients.}


\begin{thebibliography}{999}
%\bibitem{}
\setlength{\itemsep}{-0.8mm} \makeatletter
%\bibitem{} Laskin N. Fractional quantum mechanics. World Scientific; 2018.
%Stickler BA. Potential condensed-matter realization of space-fractional quantum mechanics: The onedimensional L¨¦vy crystal. Phys Rev E. 2013;88:012120.
%\small

\bibitem{b3} {\small Hilfer R. \textit{Applications of Fractional Calculus
in Physics.} World Scientific; 2000. }

\bibitem{b1} {\small Sabatier J, Agrawal OP, Machado JAT. \textit{Advances
in Fractional Calculus: Theoretical Developments and Applications in Physics
and Engineering}. Springer; 2007. }

\bibitem{b4} {\small Tarasov VE. \textit{Fractional Dynamics: Applications
of Fractional Calculus to Dynamics of Particles, Fields and Media.}
Springer; 2011. }

\bibitem{fd1} {\small Shlesinger MF, Zaslavsky GM, Klafter J. Strange
kinetics. \textit{Nature}. 1993;363:31. }

\bibitem{nG} {\small Samorodnitsky G, Taqqu MS. \textit{Stable Non-Gaussian
Random Processes.} Chapman and Hall; 1994. }

\bibitem{PR1} {\small Metzler R, Klafter J. The random walk's guide to
anomalous diffusion: a fractional dynamics approach. \textit{Phys Rep.}
2000;339:1-77. }

\bibitem{PR2} {\small Zaslavsky GM. Chaos, fractional kinetics, and
anomalous transport. \textit{Phys Rep.} 2002;371:461. }

\bibitem{Lask2} {\small Laskin N. Fractional quantum mechanics and L\'evy
path integrals. \textit{Phys Lett A.} 2000;268:298-305. }

\bibitem{Lask3} {\small Laskin N. Fractional Schr\"odinger equation. \textit{%
Phys Rev E.} 2002;66:056108.}

\bibitem{GuoXu} {\small Guo X, Xu M. Some physical applications of
fractional Schr\"odinger equation. \textit{J Math Phys.} 2006;47:082104. }

\bibitem{fd2} {\small Petroni NC, Pusterla M. L\'evy processes and
Schr\"odinger equation. \textit{Physica A.} 2009;388:824. }

\bibitem{St13} {\small Stickler BA. Potential condensed-matter realization
of space-fractional quantum mechanics: the one-dimensional L\'evy crystal.
\textit{Phys Rev E.} 2013;88:012120. }

\bibitem{Lask4} {\small Laskin N. \textit{Fractional Quantum Mechanics.}
World Scientific; 2018. }

\bibitem{fd-na} {\small Barthelemy P, Bertolotti J, Wiersma DS. A L\'evy
flight for light.\textit{\ Nature.} 2008;453:495-8. }

\bibitem{fd-np} {\small Mercadier N, Guerin W, Chevrollier M, Kaiser R.
L\'evy flights of photons in hot atomic vapours. \textit{Nat Phys.}
2009;5:602-605. }

\bibitem{Longhi} {\small Longhi S. Fractional Schr\"odinger equation in
optics. \textit{Opt Lett.} 2015;40:1117-1120. }

\bibitem{liu23} {\small Liu S, Zhang Y, Malomed BA, Karimi E. Experimental
realizations of the fractional Schr\"odinger equation in the temporal
domain. \textit{Nat Comm.} 2023;14:222. }

\bibitem{Zhang15} {\small Zhang Y, Liu X, Beli\'c MR, Zhong W, Zhang Y, Xiao
M. Propagation dynamics of a light beam in a fractional Schr\"odinger
equation. \textit{Phys Rev Lett.} 2015;115:180403. }

\bibitem{ZZ16} {\small Zhang Y, Zhong H, Beli\'c MR, Zhu Y, Zhong W, Zhang
Y, Christodoulides DN, Xiao M. $\mathcal{PT}$-symmetry in a fractional
Schr\"odinger equation.\textit{\ Laser Photonics Rev. } 2016;10:526-531. }

\bibitem{Shilong} {\small Liu S, Zhang Y, Malomed BA, Karimi E. Experimental
realizations of the fractional Schr\"odinger equation in the temporal domain.%
\textit{\ Nat Comm.} 2023;14:222. }

\bibitem{b2} {\small Monje CA, Chen YQ, Vinagre BM, Xue D, Feliu V. \textit{%
Fractional-order systems and controls.} Springer; 2010. }

\bibitem{bec-fd} {\small Stephanovich VA, Kirichenko EV, Engel G, Sinner A.
Spin-orbit-coupled fractional oscillators and trapped Bose-Einstein
condensates. \textit{Phys Rev E.} 2024;109:014222. }

\bibitem{b5} {\small Uchaikin V, Sibatov R. \textit{Fractional kinetics in
solids: Anomalous charge transport in semiconductors, dielectrics and
nanosystems.} World Scientific; 2013. }

\bibitem{book1} {\small Oldham KB, Spanier J. \textit{The Fractional
Calculus: Theory and Applications of Differentiation and Integration to
Arbitrary Order.} Academic Press; 1974. }

\bibitem{fc-rev} {\small Lovoie JL, Osler TJ, Tremblay R. Fractional
derivatives and special functions. \textit{SIAM Rev.} 1976;18:240-268. }

\bibitem{book2} {\small Kilbas A, Srivastava HM, Trujillo JJ. \textit{Theory
and Applications of Fractional Differential Equations.} Elsevier; 2006. }

\bibitem{book3} {\small Li C, Cai M. \textit{Theory and Numerical
Approximations of Fractional Integrals and Derivatives.} SIAM; 2019. }

\bibitem{caputo} {\small Caputo M. Linear models of dissipation whose $Q$ is
almost frequency independent-II. \textit{Geophys J Int.} 1967;13:529.}

\bibitem{Uchaikin} {\small Uchaikin VV. \textit{Fractional Derivatives for
Physicists and Engineers.} Springer; 2013. }

\bibitem{Riesz} {\small Cai M, Li C. On Riesz derivative. \textit{Fractional
Cal Appl Anal.} 2019;22:287-301. }

\bibitem{Benoit} {\small Mandelbrot BB. \textit{The Fractal Geometry of
Nature.} W. H. Freeman; 1982. }

\bibitem{Klein-Stoilov} {\small Klein C, Stoilov N. Multidomain spectral
approach to rational-order fractional derivatives. Stud Appl Math.
2024;152:1110-1132.}

\bibitem{KA} {\small Kivshar YS, Agrawal GP. \textit{Optical Solitons: From
Fibers to Photonic Crystals.} Academic Press; 2003. }

{\small
%\bibitem{Guo06} Guo X, Xu M. Some physical applications of fractional Schr\"odinger equation.{\it J Math Phys.} 2006;47:082104.
}

\bibitem{Huang08} {\small Huang C, Deng H, Zhang W, Ye F, Dong L.
Fundamental solitons in the nonlinear fractional Schr\"odinger equation with
a $\mathcal{PT}$-symmetric potential.\textit{\ EPL.} 2008;122:24002. }

\bibitem{Huang16} {\small Huang C, Dong L. Gap solitons in the nonlinear
fractional Schr\"odinger equation with an optical lattice. \textit{Opt Lett.}
2016;41:5636-5639. }

\bibitem{Zhang16} {\small Zhang L, Li C, Zhong H, Xu C, Lei D, Li Y, Fan D.
Propagation dynamics of super-Gaussian beams in fractional Schr\"odinger
equation: from linear to nonlinear regimes. \textit{Opt Exp.}
2016;24:14406-14418. }

\bibitem{Yao18} {\small Yao X, Liu X. Off-site and on-site vortex solitons
in space-fractional photonic lattices.\textit{\ Opt Lett.}
2018;43:5749-5752. }

\bibitem{Guo18} {\small Chen M, Zeng S, Lu D, Hu W, Guo Q. Optical solitons,
self-focusing, and wave collapse in a space-fractional Schr\"odinger
equation with a Kerr-type nonlinearity. \textit{Phys Rev E. }
2018;98:022211. }

\bibitem{Xie19} {\small Xie J, Zhu X, He Y. Vector solitons in nonlinear
fractional Schr\"odinger equations with parity-time-symmetric optical
lattices. \textit{Nonlinear Dyn.} 2019;97:1287. }

\bibitem{Zeng20} {\small Zeng L, Zeng J. Preventing critical collapse of
higher-order solitons by tailoring unconventional optical diffraction and
nonlinearities. \textit{Commun Phys.} 2020;3:26. }

\bibitem{Qiu20} {\small Qiu Y, Malomed BA, Mihalache D, Zhu X, Peng X, He Y.
Stabilization of single- and multi-peak solitons in the fractional nonlinear
Schr\"odinger equation with a trapping potential. \textit{Chaos Solitons \&
Fractals.} 2020;140:110222. }

\bibitem{Li20} {\small Li P, Dai C. Double loops and pitchfork symmetry
breaking bifurcations of optical solitons in nonlinear fractional
Schr\"odinger equation with competing cubic-quintic nonlinearities. \textit{%
Ann Phys.} 2020;532:2000048. }

\bibitem{DW} {\small Kumar S, Li P, Malomed BA. Domain walls in fractional
media. \textit{Phys Rev E.} 2022;106:054207. }

\bibitem{Zhong23} {\small Zhong M, Wang L, Li P, Yan Z. Spontaneous symmetry
breaking and ghost states supported by the fractional $\mathcal{PT}$%
-symmetric saturable nonlinear Schr\"odinger equation. \textit{Chaos.}
2023;33:013106. }

\bibitem{Zhong23prsa}  {\small  Zhong M, Yan Z. Formation of multi-peak gap solitons andstable excitations for
double-L\'evy-index andmixed fractional NLS equations with optical lattice potentials.\textit{Proc R Soc A.}
2023;479:20230222. }

\bibitem{Zhong23cp}  {\small  Zhong M, Yan Z. Spontaneous symmetry breaking and ghost statesin two-dimensional
fractional nonlinear media withnon-Hermitian potential. {\it Commun Phys.} 2023;6:92.}

\bibitem{Zhong24} {\small Zhong M, Chen Y, Yan Z, Malomed BA. Suppression of
soliton collapses, modulational instability and rogue-wave excitation in
two-L\'evy-index fractional Kerr media. \textit{Proc R Soc A.}
2024;480:20230765. }

\bibitem{Zan24} {\small Zangmo T, Mayteevarunyoo T, Malomed BA. Interactions
between fractional solitons in bimodal fiber cavities. \textit{Stud Appl
Math.} 2024;e12706. }

\bibitem{Feng-Su} {\small Feng Z, Su Y. Ground state solutions of fractional
equations with Coulomb potential and critical exponent. \textit{Stud Appl
Math.} 2024;e127236.}

\bibitem{review} {\small Malomed BA. Optical solitons and vortices in
fractional media: A mini-review of recent results. \textit{Photonics.}
2021;8:353. }

\bibitem{review2} {\small Malomed BA. Basic fractional nonlinear-wave models
and solitons. \textit{Chaos.} 2024;34:022102. }

\bibitem{rew3} {\small Kevrekidis PG, Cuevas-Maraver J, eds. \textit{%
Fractional Dispersive Models and Applications: Recent Developments and
Future Perspectives}. Springer;2024. }

\bibitem{chi2} {\small Li P, Sakaguchi H, Zeng L, Zhu X, Mihalache D,
Malomed BA. Second-harmonic generation in the system with fractional
diffraction. \textit{Chaos, Solitons \& Fractals.} 2023;173:113701. }

\bibitem{mathematical} {\small Ciaurri O, Roncal L, Stinga PR, Torrea JL,
Varona JL. Nonlocal discrete diffusion equations and the fractional discrete
Laplacian, regularity and applications. \textit{\ Adv Math.}
2018;330:688-738. }

\bibitem{Molina} {\small Molina MI. The fractional discrete nonlinear
Schr\"odinger equation. \textit{Phys Lett A.} 2020;384:126180. }

\bibitem{Molina2} {\small Molina MI. The two-dimensional fractional discrete
nonlinear Schr\"odinger equation. \textit{Phys Lett A.} 2020;384:126835. }

\bibitem{Molina-electro} {\small Molina MI. Fractional nonlinear electrical
lattice. \textit{Phys Rev E.} 2021;104:024219. }

\bibitem{fd3} {\small Molina MI. Fractionality and $\mathcal{PT}$ symmetry
in a square lattice. \textit{Phys Rev A.} 2022;106:L040202. } {\color{red} }

\bibitem{Ki13} {\small Kirkpatrick K, Lenzmann E, Staffilani G. On the
continuum limit for discrete NLS with long-range lattice interactions.
\textit{Commun Math Phys.} 2013;317:563-591.}

\bibitem{PRE95} {\small Gaididei YB, Mingaleev SF, Christiansen PL,
Rasmussen K. Effects of nonlocal dispersive interactions on self-trapping
excitations. \textit{Phys Rev E.} 1997;55:6141. }

\bibitem{CMP17} {\small Jenkinson M, Weinstein MI. Discrete solitary waves
in systems with nonlocal interactions and the Peierls-Nabarro barrier.
\textit{Commun Math Phys.} 2017;351:45-94. }

\bibitem{we} {\small Zhong M, Malomed BA, Yan Z. Dynamics of discrete
solitons in the fractional discrete nonlinear Schr\"odinger equation with
the quasi-Riesz derivative. \textit{Phys Rev E.} 2024;110:014215. }

\bibitem{MI0} {\small Benjamin TB, Feir JE. The disintegration of wave
trains on deep water part 1. \textit{J Fluid Mech.} 1967;27:417. }

\bibitem{MI1} {\small Christodoulides, DN, Joseph, RI. Discrete
self-focusing in nonlinear arrays of coupled waveguides. \textit{Opt Lett.}
1998;13,794-796. }

\bibitem{MI2} {\small Kivshar YS, Peyrard M. Modulational instabilities in
discrete lattices. \textit{Phys Rev A.} 1992;46,3198. }

\bibitem{MI3} {\small Zakharov VE, Ostrovsky LA. Modulation instability: The
beginning. \textit{Physica D}. 2009;238:540-548. }

\bibitem{Wh65} {\small Whitham GB. Non-linear dispersive waves. \textit{Proc
R Soc A.} 1965;283,238-261. }

\bibitem{Be66} {\small Bespalov VI, Talanov VI. Filamentary structure of
light beams in nonlinear liquids. \textit{Sov Phys JETP.} 1966;3:307. }

\bibitem{Os67} {\small Ostrovskii LA. Propagation of wave packets and
space-time self-focusing in a nonlinear medium. \textit{Sov Phys JETP.}
1967;24,797-800. }

\bibitem{Wa68} {\small Taniuti T, Washimi H. Self-trapping and instability
of hydromagnetic waves along the magnetic field in a cold plasma. \textit{%
Phys Rev Lett.} 1968;21,209. }

\bibitem{Ha70} {\small Hasegawa A. Observation of self-trapping instability
of a plasma cyclotron wave in a computer experiment. \textit{Phys Rev Lett.}
1970;24,1165. }

\bibitem{So07} {\small Solli DR, Ropers C, Koonath P, Jalali B. Optical
rogue waves. \textit{\ Nature.} 2007;450:1054. }

\bibitem{Kwok} {\small Chan NN, Chow KW, Kedziora DJ, Grimshaw RHJ, Ding E.
Rogue wave modes for a derivative nonlinear Schr\"{o}dinger model. \textit{%
Phys Rev E} 2014;89:03294.}

\bibitem{orw} {\small Draper L. Freak ocean waves. \textit{Oceanus.}
1964;10:13-15. }

\bibitem{Ch11} {\small Chabchoub A, Hoffmann NP, Akhmediev N. Rogue wave
observation in a water wave tank. \textit{Phys Rev Lett. } 2011;106:204502. }

\bibitem{superfluid} {\small Ganshin AN, Efimov VB, Kolmakov GV,
Mezhov-Deglin LP, McClintock PVE. Observation of an inverse energy cascade
in developed acoustic turbulence in superfluid Helium. \textit{Phys Rev
Lett. } 2008;101:065303. }

{\small
%\bibitem{Langmur} Moslem WM. 2011 Langmuir rogue waves in electron-positron
%plasmas. \textit{ Phys. Plasmas} \textbf{18}, 032301.
}

\bibitem{RW_plasma} {\small Bailung H, Sharma SK, Nakamura Y. Observation of
Peregrine solitons in a multicomponent plasma with negative ions. \textit{%
Phys Rev Lett.} 2011;107:255005. }

\bibitem{bec-rw} {\small Bludov YV, Konotop VV, Akhmediev N. Matter rogue
waves. \textit{Phys Rev A.} 2009;80:033610. }

\bibitem{bec-rw2} {\small Yan Z, Konotop VV, Akhmediev N. Three-dimensional
rogue waves in nonstationary parabolic potentials. \textit{Phys Rev A.}
2010;82:036610. }

{\small
%\bibitem{Bludov2010} Y. V. Bludov, V. V. Konotop, and N. Akhmediev, Vector rogue waves in binary mixtures of Bose-Einstein condensates, Eur. Phys. J. Special Topics {\bf 185}, 169 (2010).
}

{\small
%\bibitem{ap} Stenflo L,  Marklund M. 2010 Rogue waves in the atmosphere.\textit{ J.Plasma Phys.} \textbf{76}, 293.
}

\bibitem{Iafrati} {\small Iafrati A, Babanin A, Onorato M. Modulational
instability, wave breaking, and formation of large-scale dipoles in the
atmosphere. \textit{\ Phys Rev Lett.} 2013;110:184504. }

\bibitem{yanfrw} {\small Yan Z. Financial rogue waves. \textit{Commun Theor
Phys.} 2010;54:947. }

\bibitem{yanfrw2} {\small Yan Z. Vector financial rogue waves. \textit{Phys
Lett A}. 2011;375:4274. }

{\small
%\bibitem{St02} Strecker KE, Partridge GB, Truscott AG, Hulet RG. Formation and propagation of matter-wave soliton trains. \textit{Nature.} 2002;417,150-153.
}

{\small
%\bibitem{Ni07} Nicolin AI, Carretero-Gonz\'alez R, Kevrekidis PG. Faraday waves in Bose-Einstein condensates. \textit{Phys Rev A.} 2007;76,063609.
}

\bibitem{Ak09} {\small Akhmediev N, Ankiewicz A, Taki M. Waves that appear
from nowhere and disappear without a trace. \textit{Phys Lett A.}
2009;373:675-678. }

{\small
%\bibitem{Ha04} Haver S. A possible freak wave event measured at the Draupner jacket January 1 1995. In: \textit{Proceedings of Rogue Waves}; 2004.
}

{\small
%\bibitem{Ge22} Gemmrich J, Cicon L. Generation mechanism and prediction of an observed extreme rogue wave. \textit{Sci Rep.} 2022;12:1718.
}

\bibitem{Pe83} {\small Peregrine DH. Water waves, nonlinear Schr\"odinger
equations and their solutions. \textit{J Aust Math Soc B.} 1983;25:16. }

\bibitem{Ki10} {\small Kibler B, Fatome J, Finot C, Millot G, Dias F, Genty
G, Akhmediev N, Dudley JM. The Peregrine soliton in nonlinear fibre optics.
\textit{Nat Phys.} 2010;6:790. }

{\small
%\bibitem{Kh09} Kharif C, Pelinovsky E, Slunyaev A. \textit{Rogue Waves in the Ocean.}  Springer; 2009.
}

{\small
%\bibitem{On13} Onorato M, Residori S, Bortolozzo U, Montina A, Arecchi FT. Rogue waves and their generating mechanisms in different physical contexts. \textit{Phys Rep.} 2013;528:47.
}

{\small
%\bibitem{Ch11} Chabchoub A, Hoffmann NP, Akhmediev N. Rogue wave observation in a water wave tank. \textit{Phys Rev Lett.} 2011;106:204502.
}

{\small
%\bibitem{Bl09} Bludov YV, Konotop VV, Akhmediev N. Matter rogue waves. \textit{Phys Rev A.} 2009;80:033610.
}

{\small
%\bibitem{Ba11} Bailung H, Sharma SK, Nakamura Y. Observation of Peregrine solitons in a multicomponent plasma with negative ions. \textit{Phys Rev Lett.} 2011;107:255005.
}

{\small
%\bibitem{So07} Solli DR, Ropers C, Koonath P, Jalali B. Optical rogue waves. \textit{Nature.} 2007;450:1054.
}

{\small
%\bibitem{Ki10} Kibler B, Fatome J, Finot C, Millot G, Dias F, Genty G, Akhmediev N, Dudley JM. The peregrine soliton in nonlinear fibre optics. \textit{Nat Phys.} 2010;6:790.
}

\bibitem{RW-rew3} {\small Kharif C, Pelinovsky EN, Physical mechanisms of
the rogue wave phenomenon. \textit{Euro J Mech-B/Fluids.} 2003; 22:603-634. }

\bibitem{RW-rew4} {\small Onorato M, Residori S, Bortolozzo U, Montina A,
Arecchi FT. Rogue waves and their generating mechanisms in different
physical contexts. \textit{Phys Rep.} 2013; 528: 47. }

{\small
%\bibitem{RW-rew5} Akhmediev N, {\it et al.} Roadmap on optical rogue waves and extreme events. {\it J Opt.} 2016; 18: 063001.
}

\bibitem{RW-book} {\small Guo B, Tian L, Yan Z, Ling L, Wang YF. \textit{%
Rogue waves: mathematical theory and applications in physics}. De Gruyter;
2017. }

\bibitem{RW-rew1} {\small Dudley JM, Genty G, Mussot A, Chabchoub A, Dias F.
Rogue waves and analogies in optics and oceanography. \textit{Nat Rev Phys.}
2019; 1:675-689. }

\bibitem{RW-rew2} {\small Slunyaev AV, Pelinovsky DE, Pelinovsky EN. Rogue
waves in the sea: observations, physics, and mathematics. \textit{Phys Usp}.
2023; 66:148-172. }

\bibitem{An10} {\small Ankiewicz A, Akhmediev N, Soto-Crespo JM. Discrete
rogue waves of the Ablowitz-Ladik and Hirota equations. \textit{Phys Rev E.}
2010;82:026602. }

\bibitem{An13} {\small Ankiewicz A, Devine N, \"{U}nal M, Chowdury A,
Akhmediev N. Rogue waves and other solutions of single and coupled
Ablowitz-Ladik and nonlinear Schr\"odinger equations. \textit{J Opt.}
2013;15:064008. }

\bibitem{Yan12} {\small Yan Z, Jiang D. Nonautonomous discrete rogue wave
solutions and interactions in an inhomogeneous lattice with varying
coefficients. \textit{J Math Anal Appl.} 2012; 395: 542-549. }

\bibitem{Oh14} {\small Ohta Y, Yang JK. General rogue waves in the focusing
and defocusing Ablowitz-Ladik equations. \textit{J Phys A.} 2014;47:255201. }

\bibitem{We18} {\small Wen XY, Yan Z. Modulational instability and dynamics
of multi-rogue wave solutions for the discrete Ablowitz-Ladik equation.
\textit{J Math Phys.} 2018;59:073511. }

\bibitem{Fe21} {\small Feng BF, Ling L, Zhu ZN. A focusing and defocusing
semi-discrete complex short pulse equation and its various soliton
solutions. \textit{Proc R Soc A.} 2021;477:20200853. }

{\small
%\bibitem{Pr16} Prinari B. Discrete solitons of the focusing Ablowitz-Ladik equation with nonzero boundary conditions via inverse scattering. \textit{J Math Phys.} 2016;57:083510.
}

\bibitem{Ch14} {\small Chowdury A, Ankiewicz A, Akhmediev N. Solutions of
the higher-order Manakov-type continuous and discrete equations. \textit{%
Phys Rev E.} 2014;90:012902. }

\bibitem{We16} {\small Wen XY, Yan Z, Malomed BA. Higher-order vector
discrete rogue-wave states in the coupled Ablowitz-Ladik equations: Exact
solutions and stability. \textit{Chaos.} 2016;26:123110. }

{\small
%\bibitem{Xu24} Xu T, An LC, Li M, Xu CX. N-fold Darboux transformation of the discrete PT-symmetric nonlinear Schr\"odinger equation and new soliton solutions over the nonzero background. \textit{Stud Appl Math.} 2024;152:1338-1364.
}

\bibitem{Chen24} {\small Chen J, Pelinovsky DE. Rogue waves arising on the
standing periodic waves in the Ablowitz-Ladik equation. \textit{Stud Appl
Math.} 2024;152:147-173. }

{\small
%\bibitem{Zhong24} Zhong M, Chen Y, Yan Z, Malomed BA. Suppression of soliton collapses, modulational instability and
%rogue-wave excitation in two-L\'evy-index fractional Kerr media. {\it Proc R Soc A} 2024;480:20230765.
}

{\small
%\bibitem{Liu24} Liu H, Shen J, Geng X. Riemann-Hilbert method to the Ablowitz-Ladik equation: Higher-order cases. \textit{Stud. Appl. Math.} 2024;e12748.
}

{\small
%\bibitem{dnls} Eilbeck JC, Lomdahl PS, Scott AC. The discrete selftrapping equation. \textit{Physica D.} 1985;16:318-338.
}

\bibitem{Le08} {\small Lederer F, Stegeman GI, Christodoulides DN, Assanto
G, Segev M, Silberberg Y. Discrete solitons in optics. \textit{Phys Rep.}
2008;463:1-126. }

\bibitem{Ke09} {\small Kevrekidis PG. \textit{The discrete nonlinear
Schr\"odinger equation: mathematical analysis, numerical computations and
physical perspectives}. Springer;2009. }

\bibitem{KRB} {\small Kevrekidis PG, Rasmussen KO, Bishop AR. The discrete
nonlinear Schr\"odinger equation: A survey of recent results. \textit{Int J
Mod Phys B.} 2001;15:2833. }

\bibitem{RO} {\small Kowalski K, Rembielinski J. Relativistic massless
harmonic oscillator. \textit{Phys Rev A.} 2010;81:012118. }

\bibitem{RO2} {\small L\"orinczi J, Malecki J. Spectral properties of the
massless relativistic harmonicoscillator. \textit{J Diff Equ.}
2012;253:2846. }

\bibitem{AC4} {\small Pelinovsky DE, Kevrekidis PG, Frantzeskakis DJ.
Persistence and stability of discrete vortices in nonlinear Schr\"odinger
lattices. \textit{Physica D.} 2005;212:20-53. }

\bibitem{Be13} {\small Bertola M, Tovbis A. Asymptotics of the rational
solutions to the focusing nonlinear Schr\"odinger equation. \textit{Commun
Pure Appl Math.} 2013;66:678. }

\bibitem{Ti17} {\small Tikan A, Billet C, El G, Tovbis A, Bertola M,
Sylvestre T, Gustave F, Randoux S, Genty G, Suret P, Dudley JM. Experimental
observation of Peregrine solitons in a nonlinear fiber. \textit{Phys Rev
Lett.} 2017;119:033901. }

\bibitem{Ch18} {\small Charalampidis EG, Cuevas-Maraver J, Frantzeskakis DJ,
Kevrekidis PG. Rogue waves in ultracold bosonic seas. \textit{Rom Rep Phys.}
2018;70:504. }

\bibitem{Ho10} {\small H\"ohmann R., Kuhl U., St\"ockmann HJ, Kaplan L,
Heller EJ. Observation of Anderson localization in microwave billiards.
\textit{Phys Rev Lett.} 2010;104:093901. }

{\small
%\bibitem{AC1} Mar\'in JL, Aubry S. Breathers in nonlinear lattices: Numerical calculation from the anticontinuous limit. \textit{Nonlinearity.} 1996;9:1501-1528.
}

{\small
%\bibitem{AC2} Aubry S. Breathers in nonlinear lattices: Existence, linear stability and quantization. \textit{Physica D.} 1997;103:201-250.
}

\bibitem{AC3} {\small Pelinovsky DE, Kevrekidis PG, Frantzeskakis DJ.
Stability of discrete solitons in nonlinear Schr\"odinger lattices. \textit{%
Physica D.} 2005;212:1-19. }

\bibitem{At91} {\small Atkinson K. \textit{An Introduction to Numerical
Analysis.} John Wiley \& Sons; 1991. }

\bibitem{AC} {\small MacKay RS, Aubry S. Proof of existence of breathers for
time-reversible or Hamiltonian networks of weakly coupled oscillators.
\textit{Nonlinearity.} 1994;7:1623. }

\bibitem{Kl14} {\small Klein C, Sparber C, Markowich PA. Numerical study of
fractional nonlinear Schr\"odinger equations. \textit{Proc R Soc A.}
2014;470:20140364. }

\bibitem{Fr10} {\small Frank RL, Lenzmann E. Uniqueness of non-linear ground
states for fractional Laplacians in }$R${\small , \textit{Acta Math.}
2013;210:261-318.}

%\bibitem{collapse} {\small Chen M, Zeng S, Lu D, Hu ., and Guo Q. Optical
%solitons, self-focusing, and wave collapse in a space-fractional Schr\"{o}%
%dinger equationwith a Kerr-type nonlinearity, \textit{Phys. Rev. E. }%
%2018;98:022211.}

\bibitem{Boris01} {\small Malomed BA, Kevrekidis PG. Discrete vortex
solitons. \textit{Phys Rev E.} 2001;64:026601. }

{\small
%\bibitem{DE05} Pelinovsky DE, Kevrekidis PG, Frantzeskakis DJ. Persistence and stability of discrete vortices in nonlinear Schr\"odinger lattices. \textit{Physica D.} 2005;212:20.
}

\bibitem{Zh19} {\small Zhang X, Xu X, Zheng Y, Chen Z, Liu B, Huang C,
Malomed BA, Li Y. Semidiscrete quantum droplets and vortices. \textit{Phys
Rev Lett.} 2019;123:133901. }
\end{thebibliography}
\end{document}